\documentstyle[preprint,pra,aps,epsf,tighten]{revtex}

\begin{document}
\draft
\title{Resonant processes and exciton propagation in a frozen gas}
\author{J. S. Frasier\footnote{Electronic address:
{\texttt{jsf9k@virginia.edu}}} and
V. Celli\footnote{Electronic address: {\texttt{vc@virginia.edu}}}}
\address{Department of Physics, University of Virginia,
Charlottesville, Virginia 22904}
\date{\today}
\maketitle

\begin{abstract}
We present numerical simulations of a theory of resonant processes
in a frozen gas of excited atoms interacting via dipole-dipole
potentials that vary as $r^{-3}$, where $r$ is the interatomic
separation. The simulations calculate time-dependent averages of
transition amplitudes and transition probabilities for a single
atom in a given state interacting resonantly with a uniformly
distributed random gas of atoms in a different state. The averages
are over spatial configurations of the gas atoms, which are held
fixed while the resonant interaction creates a Frenkel exciton
that can travel from atom to atom. We check that the simulations
reproduce previously known exact results when the exciton is not
allowed to propagate [Phys.\ Rev.\ A {\bf 59}, 4358 (1999)].
Further, we develop an approximation for the average transition
amplitude that compares well with the numerical results for a wide
range of values of the system parameters.
\end{abstract}


\narrowtext


\section{Introduction}
\label{introduction}

Frozen gases are a new and in some ways ideal laboratory to test
and refine our understanding of quantum theory in a complex
system. With laser technology, the translational temperature of
the gas can be lowered to the point where it can be ignored when
discussing electronic processes, and at the same time one can
manipulate and detect these processes in the gas with an
extraordinary selectivity and precision. Frozen Rydberg gases have
the added advantage that their states are well understood and
electronic processes occur on a microsecond time scale, easily
allowing for time-resolved spectroscopy. These gases provide an
excellent means of studying many-body effects, because  the
dipole-dipole interactions which are important for binary
collisions at room temperature are still dominant in the frozen
gas, but now involve many atoms that are, to a first
approximation, standing still.

Pioneering experiments on resonant processes in frozen Rydberg
gases have been carried out by Anderson {\em et
al.}\cite{anderson,anderson2} and Mourachko {\em et
al.}\cite{mourachko}. The work presented here was motivated, in
part, by the desire to understand those experiments and the
subsequent work of Lowell {\em et al.}\cite{lowell} highlighting
the dynamic aspects of these resonant, many particle systems. The
measurements of Refs.~\cite{anderson,anderson2,lowell} are of
mixtures of ${}^{85}Rb$ atoms initially prepared in the $23s$ and
$33s$ states, henceforth to be called the $s$ and $s^{\prime }$
states, respectively. There are initially $N$ atoms in the $s$
state and $N^{\prime }$ atoms in the $s^{\prime }$ state. The
transition $ss^{\prime }\rightarrow pp^{\prime } $ is monitored,
where $p$ refers to a $24p_{1/2}$ state and $p^{\prime }$ refers
to a $34p_{3/2}$ state, at and near resonance. What is measured is
$N_{p^{\prime }}/N^{\prime }$, where $N_{p^{\prime }}$ is the
number of atoms in the $p^{\prime }$ state after an elapsed time
$t$. (Measuring $N_{p}/N$ gives nothing really new, since
$N_{p}=N_{p^{\prime }}$ at all times.) Some of the experimental
features we wish to understand are the rapid rise followed by slow
approach to saturation of the signal and the width of the
resonance lineshape (in the detuning $\Delta =\epsilon
_{p}^{\prime }+\epsilon _{p}-\epsilon _{s}^{\prime }-\epsilon
_{s}$) as a function of time. The problem is not trivial because
an initial $ss^{\prime }\rightarrow pp^{\prime }$ transition is
followed by the propagation of $sp\rightarrow ps$ and $s^{\prime
}p^{\prime }\rightarrow p^{\prime }s^{\prime }$ fluctuations,
which are responsible for the broadening of the resonance.
Although we address specifically the Rb experiments, the theory we
present here is broadly applicable to analogous processes in any
frozen gas of excited atoms.

In an earlier paper\cite{fcb} we developed the general formalism,
obtained exact results for the case when resonance broadening is
negligible, and introduced simple approximations for the
broadening. The same formalism and notation are used in this
paper. However, in Ref.~\cite{fcb} we introduced the down and up
states of an effective spin to describe the states $s$ and $p$,
and similarly for $s^{\prime }$ and $p^{\prime }$, but in this
paper we prefer to describe an $s\rightarrow p$ transition as the
creation of a Frenkel exciton\cite{bassani,knox,dexter}, which
then propagates through the medium. In this language, the initial
creation of a $pp^{\prime }$ pair launches two excitons, of types
$sp$ and $s^{\prime }p^{\prime }$, but we consider only the limit
where the propagation of the $s^{\prime}p^{\prime }$ exciton is
relatively unimportant and can be neglected. We prefer the exciton
language because the literature on excitons in solids and liquids
generally considers the dipole-dipole forces to be of major
importance, as they are in a frozen Rydberg gas, and this
literature has been useful to us in developing approximate
analytical treatments. On the other hand, in spin systems the
dominant role is usually played by the short-ranged exchange
interaction and dipole-dipole forces are given comparatively
little attention.

In this paper we present the results of numerical simulations for
the case where $N^{\prime }=1$ and $N$ is large, which corresponds
to the situation where a single $s^{\prime }$ atom is initially
present and a single $sp$ exciton is launched by the resonant
$ss^{\prime }\rightarrow pp^{\prime }$ transition. The calculated
$N_{p^{\prime }}$ can be directly compared to the experimentally
determined $N_{p^{\prime }}/N^{\prime }$ in the limit $N^{\prime
}\ll N$. When exciton propagation is suppressed, exact results are
obtained by the methods of Ref.~\cite{fcb} and serve as a check on
the numerical simulations.

We also develop analytical approximations for the average
amplitude of a resonant transition and compare them with the
results of the simulations. We find that the Cayley tree
approximation agrees well with the simulations for a broad range
of conditions, and is much better than the simple Lorentzian
approximation introduced in Ref.~\cite{fcb}. The details of the
numerical simulations are briefly outlined in
Section~\ref{simulation}. Analytical approximations for the
transition amplitude and for the exciton spectral density are
developed in Section~\ref{analytical}. Remarkably, explicit
formulas are obtained in the Cayley tree approximation.
Representative graphs of the results, both numerical and
analytical, are shown and discussed in Section~\ref{results},
which is followed by the Conclusions.

\section{DESCRIPTION OF NUMERICAL SIMULATIONS}
\label{simulation}

Before discussing how the simulations are carried out, we review
the basic elements of the theory presented in Ref.~\cite{fcb}. We
consider one atom at the origin, initially in the state
$s^{\prime}$, in interaction with a gas of $s$ atoms through a
resonant $ss^{\prime}\rightarrow pp^{\prime}$ dipole-dipole
interaction $V$. We also allow for interaction among the unprimed
atoms via an $sp\rightarrow ps$ process mediated by another
dipole-dipole potential $U$. This system is described by
Eqs.~(24a) and (24b) of Ref.~\cite{fcb}, which are
\newcounter{tlet}
\setcounter{tlet}{1}
\renewcommand{\theequation}{\arabic{equation}\alph{tlet}}
\begin{eqnarray}
\label{dilute-eqns}
 i~\dot{a}_{0} &=& \Delta a_{0} + \sum_{k=1}^{N}V_{k}c_{k}, \\
\stepcounter{tlet}\addtocounter{equation}{-1}\label{a0dot}
 i~\dot{c}_{k} &=& V_{k}a_{0} + \sum_{l=1}^{N}U_{kl}c_{l}.
\label{ckdot}
\end{eqnarray}
\renewcommand{\theequation}{\arabic{equation}}
Here $a_{0}(t)$ is the amplitude of the state in which the atom at
the origin is in state $s^{\prime }$ and all other atoms are in
state $s$, while $c_{k}(t)$ is the amplitude of the state in which
the atom at the origin is in state $p^{\prime }$ and the atom at
${\bf r}_{k}$ is in state $p$, while all the others remain in
state $s$. The quantities $V_{k}$ and $U_{kl}$ are the interaction
potentials and $\Delta=\epsilon_{p}^{\prime} +
\epsilon_{p}-\epsilon_{s}^{\prime}-\epsilon_{s}$ is the detuning
from resonance. The potential $V_{k}$ is of the dipole-dipole form
\begin{equation}
V_{k}=-\frac{2\mu \mu^{\prime }}{r_{k}^{3}} P_{2}\left(\cos\theta_{k}\right),
\end{equation}
where $\theta_{k}$ is the angle between ${\mathbf{r}}_{k}$ and the
magnetic field in the MOT used to confine the gas.\footnote{The
$P_{2}\left(\cos\theta\right)$ angular dependence is also used,
for instance, in Ref.~\cite{santos}. In the absence of a magnetic
field, each atom is in a multiplet of states and the angular
dependences of each $V_{k}$ and $U_{kl}$ are described by
matrices.} Similarly $U_{kl}$ is of the form
\begin{equation}
U_{kl}=-\frac{2\mu^{2}}{r_{kl}^{3}}
P_{2}\left(\cos\theta_{kl}\right),
\end{equation}
for $l\neq k$. We define $U_{ll} = 0$, so that the restriction
$l\neq k$ is automatic in sums of the type appearing in Eq.~(1b).
Here $\mu $ is the dipole matrix element connecting the $s$ and
$p$ states, and $\mu^{\prime}$ is the dipole matrix element
connecting the $s^{\prime}$ and $p^{\prime}$ states. The atoms are
assumed to be sufficiently cold that during the time scale of
interest they move only a very small fraction of their separation,
and therefore $V_{k}$ and $U_{kl}$ can be taken to be independent
of time.

If we consider a column vector $C\left( t\right) = \left(
a_{0}\left( t\right), c_{1}\left( t\right), c_{2}\left(
t\right),\ldots , c_{N}\left( t\right) \right)^{T}$, then clearly
we can construct a matrix $H$ such that the set of $N + 1$ coupled
differential equations (1) can be written in the form $i~\dot{C} =
HC$. If we let $\left\{ \lambda_{m}\right\}$ and $\left\{
\theta_{m}\right\}$ be the eigenvalues and eigenvectors
(represented as column vectors) of the real Hermitian matrix $H$,
respectively, and if we further define $e_{n}$ to be the column
vector of length $N+1$ with the $n$th entry set equal to one and
all others set equal to zero, then we can write $e_{n} =
\sum_{m=0}^{N}\beta_{nm}\theta_{m}$ and $\theta_{m} =
\sum_{n=0}^{N}\beta_{mn}e_{n}$, where $\beta_{nm} =
\theta_{m}^{T}\cdot e_{n}$. The system starts off in the state
described by $e_{0}$, so it follows that
\begin{eqnarray}
C\left( t\right) &=& \exp\left( -iHt\right)e_{0} \nonumber \\
&=&
\sum_{m=0}^{N}\beta_{0m}\exp\left(-i\lambda_{m}t\right)\theta_{m}
\nonumber \\
&=&
\sum_{m=0}^{N}\sum_{n=0}^{N}\beta_{0m}\beta_{mn} \exp\left(
-i\lambda_{m}t\right) e_{n}.
\end{eqnarray}
The quantity $a_{0}\left( t\right)$ is just the $e_{0}$ component
of this expression, so
\begin{equation}
a_{0}\left( t\right) = e_{0}^{T}\cdot C\left( t\right) =
\sum_{m=0}^{N} \beta_{0m}\beta_{m0}\exp\left(
-i\lambda_{m}t\right). \label{azeroeqn}
\end{equation}

Thus we see that, in this form, solving for $a_{0}$ at a given
time becomes just a matter of diagonalizing the matrix $H$. To
carry out the numerical simulations, we first generate a set of
random positions for the unprimed atoms. For the results given in
this paper these random positions are uniformly distributed, but
it is just as easy to compute for a gas blob with a Gaussian
density profile, as is the case in the
experiments\cite{anderson,anderson2,lowell}.\footnote{Results that
directly compare calculations and experiments will be reported
separately.} From these positions the potentials $V_{k}$ and
$U_{kl}$ and the matrix $H$ are generated. We then diagonalize $H$
and construct the quantity $a_{0}$ at a set of values of $t$ using
Eq.~(\ref{azeroeqn}). We do this for the same set of times for
many random distributions of the unprimed atoms and thereby obtain
an average of the quantities $a_{0}$ and $\left| a_{0}\right|^{2}$
at a given set of times over many configurations of the system.
The signal measured by the experiments is then proportional to the
average of $1 - \left| a_{0}\right|^{2}$. One can directly
generate Fourier transforms from Eq.~(\ref{azeroeqn}), as we do in
Section~\ref{results}. We can also obtain plots of the linewidth
as a function of time by generating signal versus time curves for
many values of the detuning, $\Delta$, and then determining for
each time value which of these signals is equal to one half the
value of the resonance signal.

\section{DESCRIPTION OF ANALYTICAL APPROXIMATION}
\label{analytical}

\subsection{Justification}

In developing approximations, it is useful to cast the problem in
the standard Green's function language\cite{economou}, as
discussed in Section 3 of Ref.~\cite{cellifras}. The Green's
function of Eq.~(1b) satisfies the equation
\begin{equation}
\omega G_{ln} = \delta_{ln} + \sum_{m=1}^{N}U_{lm}G_{mn},
\end{equation}
and then
\begin{equation}
a_{0}\left(\omega\right) = \frac{i}{\omega - \Delta -
\sum_{lm}V_{l}G_{lm}V_{m}}. \label{gfv}
\end{equation}
Actually, $-ia_{0}$ is the $00$ element of the Green's function
for the entire set of $N+1$ equations. Analogously, then, the
on-site Green's function for Eq.~(1b) can be written in the form
\begin{equation}
G_{nn} = \frac{1}{\omega - \sum_{lm}U_{nl}G_{lm,[n]}U_{mn}},
\label{gfu}
\end{equation}
where $G_{lm,[n]}$ is the Green's function for a medium in which
the $n$ atom is absent. It is important to keep in mind that these
Green's function equations are exact.

The Cayley approximation \cite{abou,lw} keeps only the diagonal
$\left(l = m\right)$ terms in the sum over $l$ and $m$ in
Eqs.~(\ref{gfv}) and (\ref{gfu}). Further, it assumes that each
$G_{ll,[n]}$ is independently distributed, and that there are
sufficiently many sites that $G_{ll,[n]}$ and $G_{ll}$ have the
same distribution.

We choose to set $\omega = i\alpha$ and work in Laplace space
instead of Fourier space because this makes many of the integrals
that follow more obviously convergent. Up to a point, the same
analysis can be carried out with the Fourier analogs, but Laplace
transforms are needed to solve the Cayley equations exactly in
part B of this Section. We also work with the inverses of these
Green's functions, setting $a_{0}\left(i\alpha\right) =
1/f_{0}\left(\alpha\right)$ and $G_{kk}\left(i\alpha\right) =
i/f_{k}\left(\alpha\right)$, so in the Cayley approximation we
have \setcounter{tlet}{1}
\renewcommand{\theequation}{\arabic{equation}\alph{tlet}}
\begin{eqnarray}
f_{0} &=& \alpha + i\Delta +
\sum_{k=1}^{N}\frac{V_{0k}^{2}}{f_{k}}, \label{fzero} \\
\stepcounter{tlet}\addtocounter{equation}{-1}
f_{k} &=& \alpha + \sum_{l=1}^{N}\frac{U_{kl}^{2}}{f_{l}}. \label{fk}
\end{eqnarray}
\renewcommand{\theequation}{\arabic{equation}}

\subsection{$U$ Process Only}

First we consider just the $U$ process and solve Eq.~(9b), which
is to say we look at the $sp$ exciton by itself. Note that $f_{k}$
is always real and positive. The distribution of $f_{k}$ is given
by
\begin{equation}
P\left(f_{k}\right) = \int\left\langle \delta\left(
f_{k}-\alpha-\sum_{l=1,}^{N}\frac{U_{kl}^{2}}{f_{l}}\right)\right\rangle
\prod_{m\neq k,m=1}^{N}P\left(f_{m}\right)\,df_{m},
\end{equation}
where for a uniform gas of volume $\Omega$,
\begin{equation}
\left\langle X\right\rangle = \frac{1}{\Omega^{N}}\int
\prod_{j=1}^{N}d^{3}r_{j}\,X.
\end{equation}
Writing the delta function as a Fourier transform, we see that
\begin{eqnarray}
P\left(f_{k}\right) &=&
\int_{-\infty}^{\infty}\frac{dq}{2\pi}\,\int \prod_{m\neq
k,m=1}^{N}P\left(f_{m}\right)df_{m}\, e^{iq\left( f_{k}-\alpha
\right)} \left\langle \exp\left(-iq\sum_{l=1}^{N}
\frac{U_{kl}^{2}}{f_{l}}\right)\right\rangle \nonumber \\ &=&
\int_{-\infty}^{\infty}\frac{dq}{2\pi}\,e^{iq\left( f_{k}-\alpha
\right)}\int \prod_{m\neq k,m=1}^{N}P\left(f_{m}\right)df_{m}\,
\frac{1}{\Omega^{N-1}}\int \prod_{l\neq k,l=1}^{N}d^{3}\,r_{l}
\exp\left(-iq\frac{U_{kl}^{2}}{f_{l}}\right).
\end{eqnarray}
Using the fact that $\left(\prod_{m}P_{m}\right)
\left(\prod_{l}E_{l}\right) = \prod_{m}P_{m}E_{m}$, we can
write
\begin{eqnarray}
P\left(f_{k}\right)&=&
\int_{-\infty}^{\infty}\frac{dq}{2\pi}\,e^{iq\left( f_{k}-\alpha
\right)}\prod_{m\neq k,m=1}^{N}\int P\left(f_{m}\right)df_{m}\,
\frac{1}{\Omega}\int d^{3}\,r_{m}
\exp\left(-iq\frac{U_{km}^{2}}{f_{m}}\right) \nonumber \\ &=&
\int_{-\infty}^{\infty}\frac{dq}{2\pi}\,e^{iq\left( f_{k}-\alpha
\right)} \left[\int_{0}^{\infty} df_{1}\,P\left(f_{1}\right)
\frac{1}{\Omega}\int d^{3}r_{1}\,
\exp\left(-iq\frac{U_{k1}^{2}}{f_{1}}\right)\right]^{N-1}
\nonumber
\\ &=& \int_{-\infty}^{\infty}\frac{dq}{2\pi}\,e^{iq\left(
f_{k}-\alpha \right)} \left(1-\int_{0}^{\infty}
df\,P\left(f\right) \frac{1}{\Omega}\int d^{3}r\,
\left\{1-\exp\left[-iq\frac{U^{2}\left(r\right)}{f}\right]
\right\}\right)^{N-1},
\end{eqnarray}
where in the last step we have simply changed variables from
${\mathbf{r}}_{1}$ to ${\mathbf{r}} = {\mathbf{r}}_{1} -
{\mathbf{r}}_{k}$. Using the averaging methods detailed in
Eqs.~(8--10) and the Appendix of Ref.~\cite{fcb}, we see that
\begin{equation}
P\left(f\right) =
\int_{-\infty}^{\infty}\frac{dq}{2\pi}\,e^{iq\left( f-\alpha
\right)} \exp\left[-u\sqrt{iq}Q\left(\alpha\right)\right],
\end{equation}
where
\begin{equation}
Q\left(\alpha\right) = \int_{0}^{\infty} \frac{df}{\sqrt{f}}\,
P\left(f\right),
\end{equation}
and
\begin{equation}
u = \frac{16\pi^{3/2}}{9\sqrt{3}}\mu^{2}\frac{N}{\Omega}.
\label{u}
\end{equation}
Using
\begin{equation}
\int_{-\infty}^{\infty}\frac{dq}{2\pi}\, e^{iqA-\sqrt{iq}B} =
\frac{1}{2\sqrt{\pi}}\frac{B}{A^{3/2}}\exp\left(-\frac{B^{2}}{4A}\right)
\theta\left(A\right),
\end{equation}
for $A$ and $B$ real, we have
\begin{equation}
P\left(f\right) =
\frac{1}{2\sqrt{\pi}}\frac{uQ\left(\alpha\right)}{\left( f-\alpha
\right)^{3/2}}
\exp\left[-\frac{u^{2}Q^{2}\left(\alpha\right)}{4\left( f-\alpha
\right)}\right] \theta\left(f-\alpha\right).
\end{equation}
and so
\begin{eqnarray}
Q\left(\alpha\right) &=& \int_{\alpha}^{\infty}
\frac{df}{\sqrt{f}}\,
\frac{1}{2\sqrt{\pi}}\frac{uQ\left(\alpha\right)}{\left(f-\alpha\right)^{3/2}}
\exp\left[-\frac{u^{2}Q^{2}\left(\alpha\right)}{4\left(f-\alpha\right)}\right]
 \nonumber \\ &=&
\frac{1}{\sqrt{\alpha}}\exp\left[\frac{u^{2}Q^{2}\left
(\alpha\right)}{4\alpha} \right]
{\mathrm{erfc}}\left[\frac{uQ\left(\alpha\right)}{2\sqrt{\alpha}}\right].
\label{Qeqn}
\end{eqnarray}
We also have
\begin{eqnarray}
\left\langle\frac{1}{f}\right\rangle &=& \int_{0}^{\infty}\frac{df}{f}\,
P\left(f\right) \nonumber \\ &=& \frac{1}{\alpha} -
\frac{\sqrt{\pi}}{2}\frac{uQ\left(\alpha\right)}{\alpha^{3/2}}
\exp\left[\frac{u^{2}Q^{2}\left(\alpha\right)}{4\alpha}\right]
{\mathrm{erfc}}\left[\frac{uQ\left(\alpha\right)}{2\sqrt{\alpha}}\right],
\end{eqnarray}
or
\begin{equation}
\left\langle\frac{1}{f}\right\rangle = \frac{1}{\alpha}\left[
1-\frac{\sqrt{\pi}}{2}uQ^{2}\left(\alpha\right) \right]. \label{fkinv}
\end{equation}
Thus we can solve numerically the transcendental relation for
$Q\left(\alpha\right)$ given in Eq.~(\ref{Qeqn}) and plug it into
Eq.~(\ref{fkinv}) to find $\left\langle 1/f\right\rangle$. It is
important to note that Eq.~(\ref{Qeqn}) defines $Q$ as an analytic
function of $\alpha$ in the half-plane where
${\mathrm{Re}}\,\alpha > 0$. Thus $\left\langle 1/f\right\rangle$
can be analytically continued, in the variable $\omega=i\alpha$,
from the positive imaginary axis to the entire upper half-plane.

\subsection{$U$ and $V$ Processes in the Absence of $\Delta$}

Now we allow the $V$ process to be present, but we still ignore
the detuning, $\Delta$. We start with the equations
\setcounter{tlet}{1}
\renewcommand{\theequation}{\arabic{equation}\alph{tlet}}
\begin{eqnarray}
f_{0} &=& \alpha + \sum_{k=1}^{N}\frac{V_{0k}^{2}}{f_{k}}, \\
\stepcounter{tlet}\addtocounter{equation}{-1}
f_{k} &=& \alpha + \sum_{l=1,l\neq k}^{N}\frac{U_{kl}^{2}}{f_{l}}.
\end{eqnarray}
\renewcommand{\theequation}{\arabic{equation}}
Repeating the analysis we did before, we have
\begin{equation}
P_{0}\left(f_{0}\right) = \int\left\langle\delta\left(f_{0} -
\alpha - \sum_{k=1}^{N}\frac{V_{0k}^{2}}{f_{k}}
\right)\right\rangle\prod_{l=1}^{N}P\left(f_{l}\right)df_{l},
\end{equation}
which leads to the form
\begin{equation}
P_{0}\left(f_{0}\right) = \int_{-\infty}^{\infty}\frac{dq}{2\pi}\,
e^{iq\left(f_{0}-\alpha\right)} \left\{\int_{0}^{\infty}df\,
P\left(f\right)\frac{1}{\Omega}\int
d^{3}r\exp\left[-iq\frac{V^{2}\left(r\right)}{f}\right]
\right\}^{N}.
\end{equation}
At this point it is clear from our previous work that this
equation leads to
\begin{equation}
P_{0}\left(f_{0}\right) = \frac{1}{2\sqrt{\pi}}
\frac{vQ\left(\alpha\right)}{\left(f_{0}-\alpha\right)^{3/2}}
\exp\left[
-\frac{v^{2}Q^{2}\left(\alpha\right)}{4\left(f_{0}-\alpha\right)}
\right]\theta\left(f_{0}-\alpha\right),
\end{equation}
where
\begin{equation}
v = \frac{16\pi^{3/2}}{9\sqrt{3}}\mu\mu^{\prime}\frac{N}{\Omega}, \label{v}
\end{equation}
and $Q\left(\alpha\right)$ is still given by Eq.~(\ref{Qeqn}), and
depends on $u$. Therefore we obtain the result
\begin{eqnarray}
\left\langle\frac{1}{f_{0}}\right\rangle &=&
\int_{0}^{\infty}\frac{df_{0}}{f_{0}}\,P_{0}\left(f_{0}\right)
\nonumber \\ &=& \frac{1}{\alpha}-\frac{\sqrt{\pi}}{2}
\frac{vQ\left(\alpha\right)}{\alpha^{3/2}}
\exp\left[\frac{v^{2}Q^{2}\left(\alpha\right)}{4\alpha}\right]
{\mathrm{erfc}}\left[ \frac{vQ\left(\alpha\right)}{2\sqrt{\alpha}}
\right], \label{fzerowithu}
\end{eqnarray}
which again can be analytically continued.

\subsection{$U$ and $V$ Processes in the Presence of $\Delta$}

If we further consider the effect of a detuning, $\Delta$, then we
have the full set of Eqs.~(9). The equation for $f_{k}$ is
unchanged from the previous case, but because the equation for
$f_{0}$ is now complex we have to consider a distribution in the
two variables $f_{0}^{R} = {\mathrm{Re}}\,f_{0}$ and $f_{0}^{I} =
{\mathrm{Im}}\,f_{0}$. We have
\begin{equation}
P_{0}\left(f_{0}^{R},f_{0}^{I}\right) =
\int\left\langle\delta\left(f_{0}^{R} - \alpha -
\sum_{k=1}^{N}\frac{V_{0k}^{2}}{f_{k}}
\right)\delta\left(f_{0}^{I} - \Delta
\right)\right\rangle\prod_{l=1}^{N}P\left(f_{l}\right)df_{l}.
\end{equation}
Clearly the delta function involving $f_{0}^{I}$ does not
participate in the averaging or in the integrations over $f_{l}$.
Therefore we can treat the part involving $f_{0}^{R}$ just as we
did before, and the $f_{0}^{I}$ part will just be carried along.
Hence we find
\begin{equation}
P_{0}\left(f_{0}^{R},f_{0}^{I}\right) = \frac{1}{2\sqrt{\pi}}
\frac{vQ\left(\alpha\right)}{\left(f_{0}^{R}-\alpha\right)^{3/2}}
\exp\left[
-\frac{v^{2}Q^{2}\left(\alpha\right)}{4\left(f_{0}^{R}-\alpha\right)}
\right]\theta\left(f_{0}^{R}-\alpha\right)\delta\left(f_{0}^{I} -
\Delta\right).
\end{equation}
It follows, then, that
\begin{eqnarray}
\left\langle\frac{1}{f_{0}}\right\rangle &=&
\int_{0}^{\infty}df_{0}^{R}\int_{-\infty}^{\infty}df_{0}^{I}\,
\frac{1}{f_{0}^{R} + i f_{0}^{I}}
P_{0}\left(f_{0}^{R},f_{0}^{I}\right) \nonumber \\ &=&
\frac{1}{\alpha + i\Delta}-\frac{\sqrt{\pi}}{2}
\frac{vQ\left(\alpha\right)}{\left(\alpha + i\Delta\right)^{3/2}}
\exp\left[ \frac{v^{2}Q^{2}\left(\alpha\right)}{4\left(\alpha +
i\Delta\right)}\right] {\mathrm{erfc}}\left[
\frac{vQ\left(\alpha\right)}{2\sqrt{\alpha + i\Delta}} \right].
\end{eqnarray}

\subsection{Other Approximations}

It is also possible to make approximations other than the one we
have chosen. For example, instead of working with Eq.~(9b), one
can instead choose to work with
\begin{equation}
f_{k} = \alpha + \sum_{l=1,}^{N}\frac{U_{kl}^{2}}{f_{k}},
\label{fhubbard}
\end{equation}
which gives explicitly
\begin{equation}
f_{k} = \frac{1}{2}\left(\alpha +
\sqrt{\alpha^{2}+4\sum_{l}U_{kl}^{2}}\right).
\end{equation}
The standard averaging procedure then leads to
\begin{equation}
\left\langle\frac{1}{f}\right\rangle =
\frac{2}{u\sqrt{\pi}} +
\frac{\alpha}{u^{2}}\left[\exp\left(\frac{u^{2}}{\alpha^{2}}\right)
{\mathrm{erfc}}\left(\frac{u}{\alpha}\right)-1\right].
\end{equation}
Alternatively, we can replace $1/f_{l}$ by its average, obtaining
\begin{equation}
f_{k} =
\alpha + \sum_{l=1}^{N}U_{kl}^{2}
\left\langle\frac{1}{f}\right\rangle,
\label{fcpa}
\end{equation}
which leads to
\begin{equation}
\alpha\left\langle\frac{1}{f}\right\rangle +
\frac{\sqrt{\pi}}{2}\frac{u}{\sqrt{\alpha}}
\sqrt{\left\langle\frac{1}{f}\right\rangle} \exp\left(\frac{u^{2}}{4\alpha}
\left\langle\frac{1}{f}\right\rangle\right)
{\mathrm{erfc}}\left(\frac{u}{2\alpha}
\sqrt{\left\langle\frac{1}{f}\right\rangle}\right) = 1.
\end{equation}

Eqs.~(\ref{fhubbard}) and (\ref{fcpa}) correspond to a Hubbard
band and a CPA approximation, respectively. We have examined these
approximations and found that they give results that do not agree
with our simulations as well as the results obtained by the Cayley
approximation. This is not at all surprising, since in the Cayley
equations (9) the on-site Green's functions for the $U$ problem
are allowed to vary from site to site, whereas in
Eqs.~(\ref{fhubbard}) and (\ref{fcpa}) they are not.

\section{RESULTS}
\label{results}

In this section we present and discuss the results of the
numerical simulations described in Section \ref{simulation} in
comparison, when possible, with exact analytical results or with
approximations that are good enough to be used instead of the
simulations. We require that the graph of an approximate result
should not entail much more numerical labor than the corresponding
graph from the numerical simulations. Further, we omit comparisons
of Laplace transforms, because two similar-looking Laplace
transforms can correspond to distinctly different results in the
time domain and to different spectral densities in the (real)
frequency domain. We found that only the Cayley tree approximation
for $\left\langle a_{0}\right\rangle$, as developed in Section
\ref{analytical}, comes close to meeting all these criteria and
can be considered as ``good'' for a broad range of conditions,
although in special limits simpler approximations suffice. We do
not have, at this time, a good approximation for $\left\langle
|a_{0}|^{2}\right\rangle $.

All the analytical results, exact and approximate, are for a
single primed atom in an ideal, infinite gas of unprimed atoms
with purely dipolar interactions. The simulations presented here
were carried out for $N$ unprimed atoms randomly distributed in a
cubic volume floating in vacuum, with a single primed atom at the
center of the cube. One can change the position of the unprimed
atom (to study surface effects), the shape of the volume, the
boundary conditions, and the interaction. One can also consider a
non-uniform gas blob, such as one with a Gaussian profile, and put
in some correlation between the positions of the atoms. Here we
focus on the uniform, ideal gas as a standard.

For compactness of presentation, all energy and inverse time
variables are scaled by the effective strength of the $ss^{\prime
}\rightarrow pp^{\prime } $ transition, $v$, defined in
Eq.~(\ref{v}). We have then results as a function of time or
frequency and of two parameters: $\Delta/v$, where $\Delta$ is the
detuning from resonance, and $u/v$, where $u$ is the equivalent of
$v$ for the $sp\rightarrow ps$ transition and is defined in
Eq.~(\ref{u}). We display results for the following values of $u$:

\noindent\textbf{\textit{u=0}}

If the potential $U$ is set equal to zero in Eq.~(1b), then
Eqs.~(1) can be solved and averaged analytically as is done in
Ref.~\cite{fcb}. This case therefore provides a convenient way to
check to see if our numerical routines are working properly. It
also provides a nice way to get a feeling for how many unprimed
atoms need to be included and over how many realizations of the
system we need to average to obtain good statistical results. In
this case, of course, the $sp$ exciton does not travel.

\noindent\textbf{\textit{u=v/4}}

We report some results for this case in order to show the gradual transition
from the non-propagating to the propagating exciton, and also because this
value of $u/v$ corresponds to the Rb case when there are, initially, a few $s
$ atoms in a gas of $s^{\prime }$ atoms ($N\ll N^{\prime }$). In effect, one
interchanges the roles of primed and unprimed atoms with respect to the case
$u=4v$ (see below), so that the $s^{\prime }p^{\prime }$ exciton is now the
one that propagates.

\noindent\textbf{\textit{u=v}}

Resonant coupling to the excitons is nearly optimal in this case.
In addition, for $\Delta =0$ finding $a_{0}$ is equivalent to
finding the on-site Green's function $G_{00}$ for a gas of $N+1$
atoms. Then $-\left(1/\pi\right) {\mathrm{Im}}\,
G_{00}\left(\omega\right)=\left(1/\pi
\right){\mathrm{Re}}\,a_{0}\left(\omega\right)$ is the spectral
density for exciton propagation. This and other quantities related
to exciton propagation can then be studied independently from the
mechanism of resonant excitation.

\noindent\textbf{\textit{u=4v}}

This value is chosen because it is close to the value $u=3.96v$
that pertains to the Rb
experiments\cite{anderson,anderson2,lowell}. It also serves to
illustrate, in general, what happens when exciton propagation is a
large effect.

\subsection{Results for $N_{p^{\prime }}/N^{\prime }$}

The four parts of Fig.~\ref{azero2} display the numerical
simulations of $1-\left\langle \left| a_{0}\right|
^{2}\right\rangle $, corresponding to the experimental signal
$N_{p^{\prime }}/N^{\prime }$, as a function of $vt$ for
$u/v=0,1/4,1$ and $4$, and enough values of the detuning $\Delta$
to give an idea of the resonance width in each case. The FWHM at
large $t$ can be directly deduced from the intercepts of the
graphs with the right edge of the figure, and is discussed more
fully in subsection D.

In addition, Fig.~1a shows the exact results for $u=0$ that were
reported in Ref.~\cite{fcb}. Comparison with the numerical
simulations, which were carried out for $N=100$, shows that the
infinite system is already well simulated, but that noticeable
wiggles remain after averaging over 10000 realizations. The reason
for this persistence of fluctuations is that most realizations do
not look at all like the average. In fact, the variance with
realization (not shown) is comparable to the average and in
particular for small $t$ the average is given by $\left(\sqrt{\pi
}/2\right)vt$, while each realization is quadratic in $t$, as
already discussed in Ref.~\cite{fcb}. It can be seen from the
graphs that the linear rise law, $\left(\sqrt{\pi}/2\right)vt$, is
independent both of $\Delta$ and of $u$ and holds for a
substantial range of $vt$.

\subsection{Results for $\left\langle a_{0}\left(t\right)\right\rangle $}

The four parts of Figs.~\ref{azero_real} and \ref{azero_imag}
display the numerical simulations of ${\mathrm{Re}}\left\langle
a_{0}\right\rangle $ and ${\mathrm{Im}}\left\langle
a_{0}\right\rangle ,$ respectively, as a function of $vt$  for
$u/v=0,1/4,1$, and $4$. Recall that $\left\langle
a_{0}\right\rangle $ is the probability amplitude that resonant
exciton creation has not occurred during the time $t$. The
striking feature of these graphs is that, off-resonance,
oscillations with an approximate period $\Delta $ persist for a
long time, not only when the exciton does not propagate ($u=0$),
but even for $u=4v$.

The graph for $u=0$ can be obtained analytically by the same
methods used in Ref.~\cite{fcb} for computing $1-\left\langle
\left| a_{0}\right| ^{2}\right\rangle $. We give here only the
following result for the Laplace transform:
\begin{equation}
\left\langle a_{0}\left(i\alpha\right)\right\rangle
=\frac{1}{\alpha+i\Delta} -
\frac{\sqrt{\pi}}{2}\frac{v}{\left(\alpha+i\Delta\right)^{3/2}
\sqrt{\alpha}}
\exp\left[\frac{v^{2}}{4\alpha\left(\alpha+i\Delta\right)}\right]
{\mathrm{erfc}}\left[\frac{v}{2\sqrt{\alpha\left(\alpha +
i\Delta\right)}}\right]. \label{a0alpha}
\end{equation}
From this, $\left\langle a_{0}\left(t\right)\right\rangle$ has
been obtained by expanding in powers of $1/\alpha$ and inverting
the Laplace transform term by term. As in Ref.~\cite{fcb}, various
formulas involving special functions can also be obtained. For
$u=0$, the exact analytical results (not shown here) agree with
the numerical simulations about as well as in Fig.~1a. The initial
dependence on time is given by
\begin{equation}
\left\langle a_{0}\right\rangle \simeq 1 - \frac{\sqrt{\pi }}{2}vt
- i\Delta t  \label{a0lin}
\end{equation}
Looking at the graphs, it seems that the initial decrease of
${\mathrm{Re}} \left\langle a_{0}\right\rangle $ depends on $u$,
but this perception only indicates that the range of validity of
the linear approximation (\ref{a0lin}) is small. On the other
hand, the graphs of $\left\langle \left| a_{0}\right|
^{2}\right\rangle $ in Fig.~\ref{azero2} look linear and
independent of $u$ over a wide range of $vt$. Note also that
$\left| \left\langle a_{0}\right\rangle \right| ^{2}\simeq
1-\sqrt{\pi }vt$ decreases initially twice as fast as
$\left\langle \left| a_{0}\right| ^{2}\right\rangle \simeq
1-(\sqrt{\pi }/2)vt$. Of course, it must be true that the (mod)
square of the average is smaller than or equal to the average of
the square, but the above result shows that $\left| \left\langle
a_{0}\right\rangle \right| ^{2}$ is not even a fair approximation
to $\left\langle \left| a_{0}\right| ^{2}\right\rangle $.

\subsection{Results for the Spectral Density}

Figs.~\ref{u=0_azero_fourier} to \ref{u=4v_azero_fourier} present
graphs of ${\mathrm{Re}}\left\langle a_{0}\left(\omega
\right)\right\rangle $ that correspond to the graphs of
$\left\langle a_{0}\left(t\right)\right\rangle $ in
Figs.~\ref{azero_real} and \ref{azero_imag}. For $u=v$ and $\Delta
=0$, ${\mathrm{Re}}\left\langle
a_{0}\left(\omega\right)\right\rangle $ is the exciton spectral
density (times $\pi$); more generally, it is an excitation
spectral density for the resonant process $ss^{\prime }\rightarrow
pp^{\prime }$. Each Figure gives a comparison of the numerical
simulation with the Cayley approximation, which in particular for
$u=0$ reduces correctly to the exact result of
Eq.~(\ref{a0alpha}), obtained from Eqs.~(\ref{fzerowithu}) and
(\ref{Qeqn}) by analytic continuation. We compare spectral
densities, rather than time graphs, in part because obtaining
$\left\langle a_{0}\left(t\right)\right\rangle $ for the Cayley
approximation is laborious, and in part because the comparison of
spectral densities is instructive, as we now discuss.

In the notation of Section \ref{simulation}, the spectral density
is obtained from Eq.~(\ref{azeroeqn}) as
\begin{equation}
{\mathrm{Re}}\,a_{0}\left( \omega +i\varepsilon \right)
={\mathrm{Im}}\sum_{m=0}^{N} \frac{\beta_{0m}\beta_{m0}}{\omega
-\lambda_{m}+i\varepsilon }
\end{equation}
in the limits $\varepsilon \rightarrow 0$ and $N\rightarrow \infty
$. In practice it is difficult to go beyond $N=1000$, and one must
choose $\varepsilon $ to be fairly large to obtain a smooth plot
for the spectral density where the eigenvalues are sparse, even
after averaging over 10000 realizations. On the other hand, the
distribution of eigenvalues for a random dipolar gas is highly
structured near $\omega =0$, and if $\varepsilon $ is too large
this structure will be missed. Rather than spending much time to
go to large $N$ and refine the numerical technique, we set
$N=100$, plot the results for $\varepsilon /v=0.1$ and $0.01$, and
let the eye extrapolate to the correct limit. It is remarkable
that the Cayley approximation agrees best with the smallest value
of $\varepsilon /v$, except for wiggles that are usually in
regions of sparse eigenvalues. This is exactly how the exact
spectral density is expected to compare with a calculation for
finite $N$. A finite $\varepsilon$ can also be viewed as a
simulation of the decoherence introduced by atomic motions, or of
losses to other channels, such as the decay of the excited atomic
states due to blackbody radiation\cite{gallagher}. Thus,
$\varepsilon$ is almost equivalent to the $\gamma$ of the
Lorentzian broadening model of Ref.~\cite{fcb}, the difference
being that in Ref.~\cite{fcb} $i\gamma $ is added to $\omega$ for
the Fourier transform of Eq.~(1b) only, while here $i\varepsilon $
is added to every $\omega$. As $\varepsilon$ increases, the highly
structured spectral densities of the ideal frozen gas become
increasingly similar to Lorentzian lines. Because of unavoidable
losses and incoherences, the experimental spectral densities will
resemble the results of the simulations for some finite
$\varepsilon$.

For $u=0$ (Fig.~\ref{u=0_azero_fourier}) the spectral density is
quite different from a Lorentzian. At $\Delta = 0$ there is a
quasi-gap around $\omega = 0$. As $\Delta$ increases the gap
extends from $\omega = 0$ to $\omega = \Delta$.

For $u=v/4$ (Fig.~\ref{u=0.25v_azero_fourier}) the sharp features
of the $u=0$ spectrum are rounded off and the gap is beginning to
be filled, but is still recognizable.

For $u=v$ (Fig.~\ref{u=v_azero_fourier}) the spectrum at $\Delta =
0$ has the appearance of a single, asymmetric line. At $\Delta =
v$ and $\Delta = 2v$ a shoulder on the left of the line is the
remnant of the gap of Fig.~\ref{u=0_azero_fourier}.

For $u=4v$ (Fig.~\ref{u=4v_azero_fourier}) there is a single line
for all values of $\Delta$, which becomes increasingly sharp as
$u$ increases. The lineshape is not Lorentzian, but even a small
additional broadening will make it nearly so, as indicated by the
calculation for $\varepsilon/v = 0.01$. This sharp line leads to
the damped sinusoidal oscillations seen in the $u=4v$ graphs of
Figs.~\ref{azero_real} and \ref{azero_imag}. The general
phenomenon of line narrowing can be discussed using the simple
Lorentzian model for the exciton band, in which Eq.~(1b) of
Ref.~\cite{fcb} is replaced by
\begin{equation}
i\dot{c_{k}} = V_{k}a_{0} - i\gamma c_{k}.
\end{equation}
Solving the equations in this limit, one finds a pole at
\begin{equation}
\omega = -i\frac{\gamma\sum_{k}V_{k}^{2}}{\gamma^{2} + \Delta^{2}}
+ \Delta,
\end{equation}
which gives a decreasing width as $u$ increases, since $\gamma$ is
proportional to $u$. One cannot simply replace $\sum_{k}V_{k}^{2}$
by its average, which does not even exist. Instead, one can
Fourier transform the average to find the time dependence
\begin{equation}
\exp\left(-i\Delta t - \sqrt{\gamma_{eq}t}\right),
\end{equation}
where
\begin{equation}
\gamma_{eq} = \frac{v^{2}\gamma}{\gamma^{2} + \Delta^{2}},
\end{equation}
and $v$ has been defined in Eq.~(\ref{v}). A similar argument was
applied in Ref.~\cite{fcb} to find the time dependence of
$\left\langle\left|a_{0}\right|^{2}\right\rangle$.

From the comparison with the simulations, it is apparent that a
shortcoming of the Cayley approximation is that it gives a
symmetric spectral density at resonance ($\Delta =0$), while the
correct spectral density is slightly asymmetric, except for $u=0$.
Correspondingly, $\left\langle a_{0}\left(t\right)\right\rangle $
is not purely real, as can be seen in Fig.~\ref{azero_imag}. For
$\Delta \neq 0$, the spectral density at $\omega < \Delta$ is
underestimated by the Cayley approximation.

Clearly, the asymmetry of the spectral density at $\Delta =0$
comes from the terms with $l\neq m$ in the denominator of
Eq.~(\ref{gfv}), which are neglected in the Cayley approximation.
It is surprising that these terms, which correspond to exciton
propagation other than back and forth on the Cayley tree, have
such small effect on $\left\langle a_{0}\right\rangle$ at $\Delta
=0$. At finite $\Delta$, the underestimate of
${\mathrm{Re}}\left\langle a_{0}\left(\omega\right)\right\rangle $
for $\omega < \Delta$ is probably due to the neglect of these same
terms. It seems complicated to include these terms, even
approximately, in the Cayley tree formulation, but we have already
shown how this can be done in the ``CPA''
approximation\cite{cellifras}.

\subsection{Results for the Resonance Linewidth}

\begin{figure}[tbp]
\begin{center}
\hbox{\epsffile{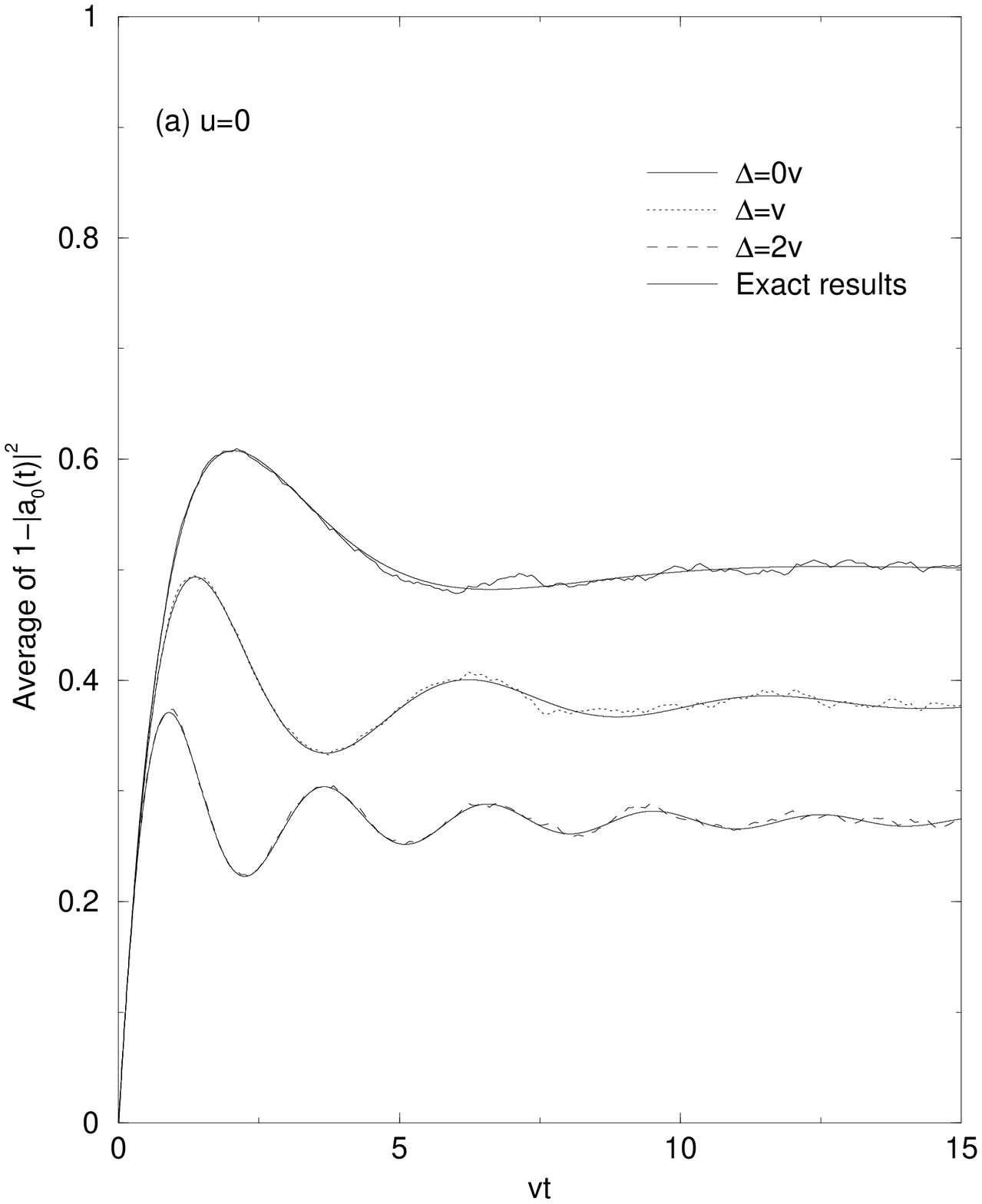}}
\pagebreak
\hbox{\epsffile{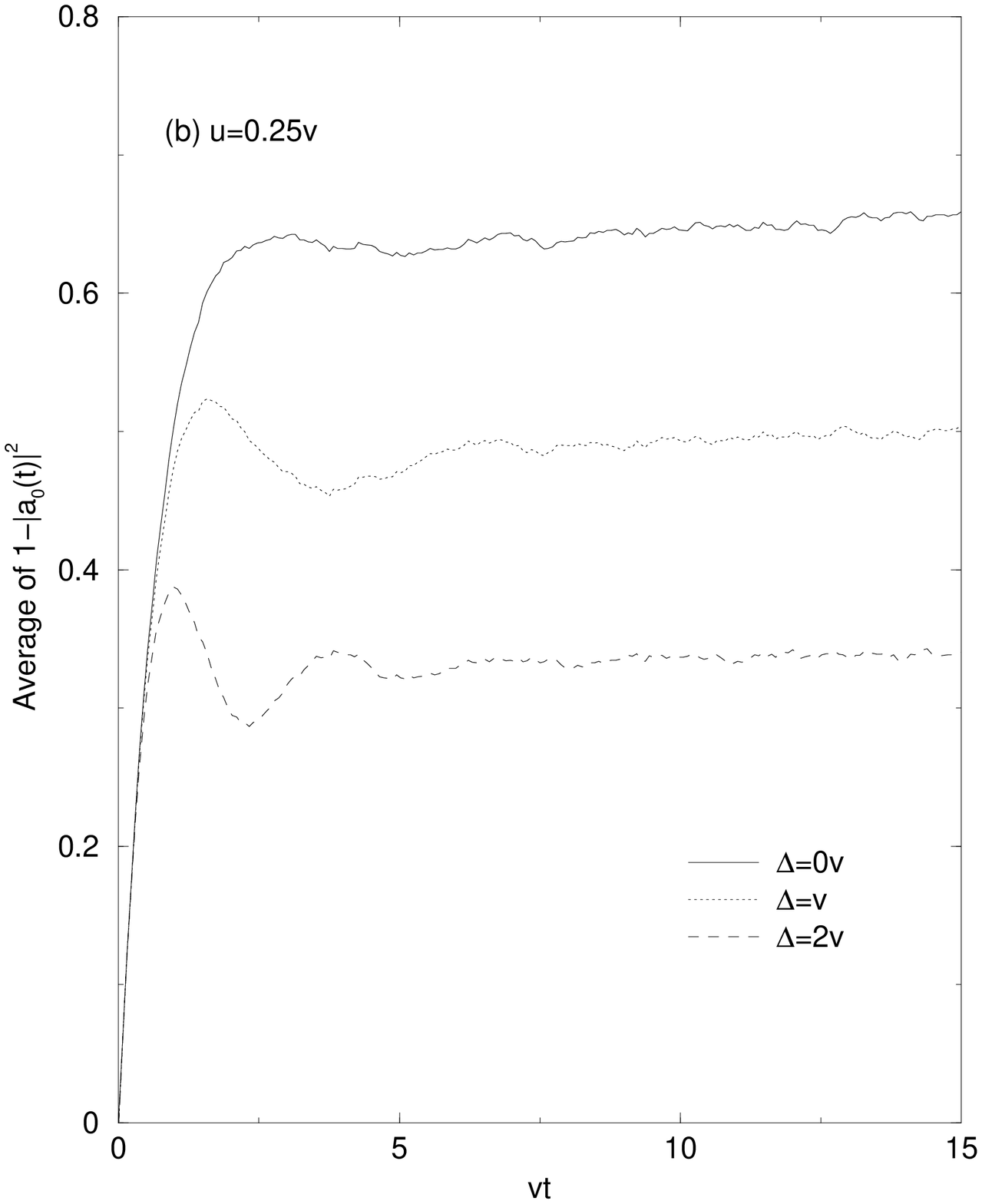}}
\pagebreak
\hbox{\epsffile{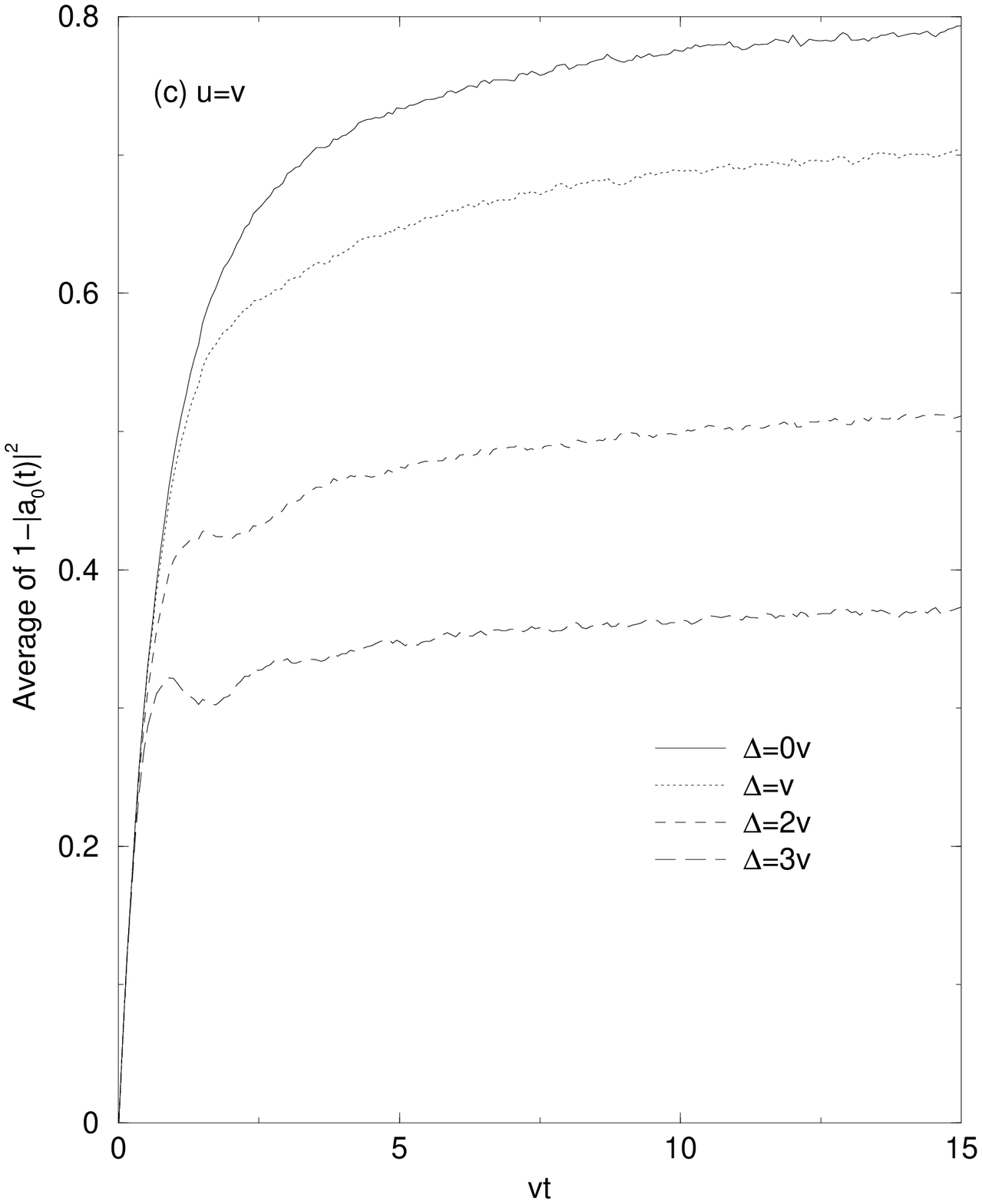}}
\pagebreak
\hbox{\epsffile{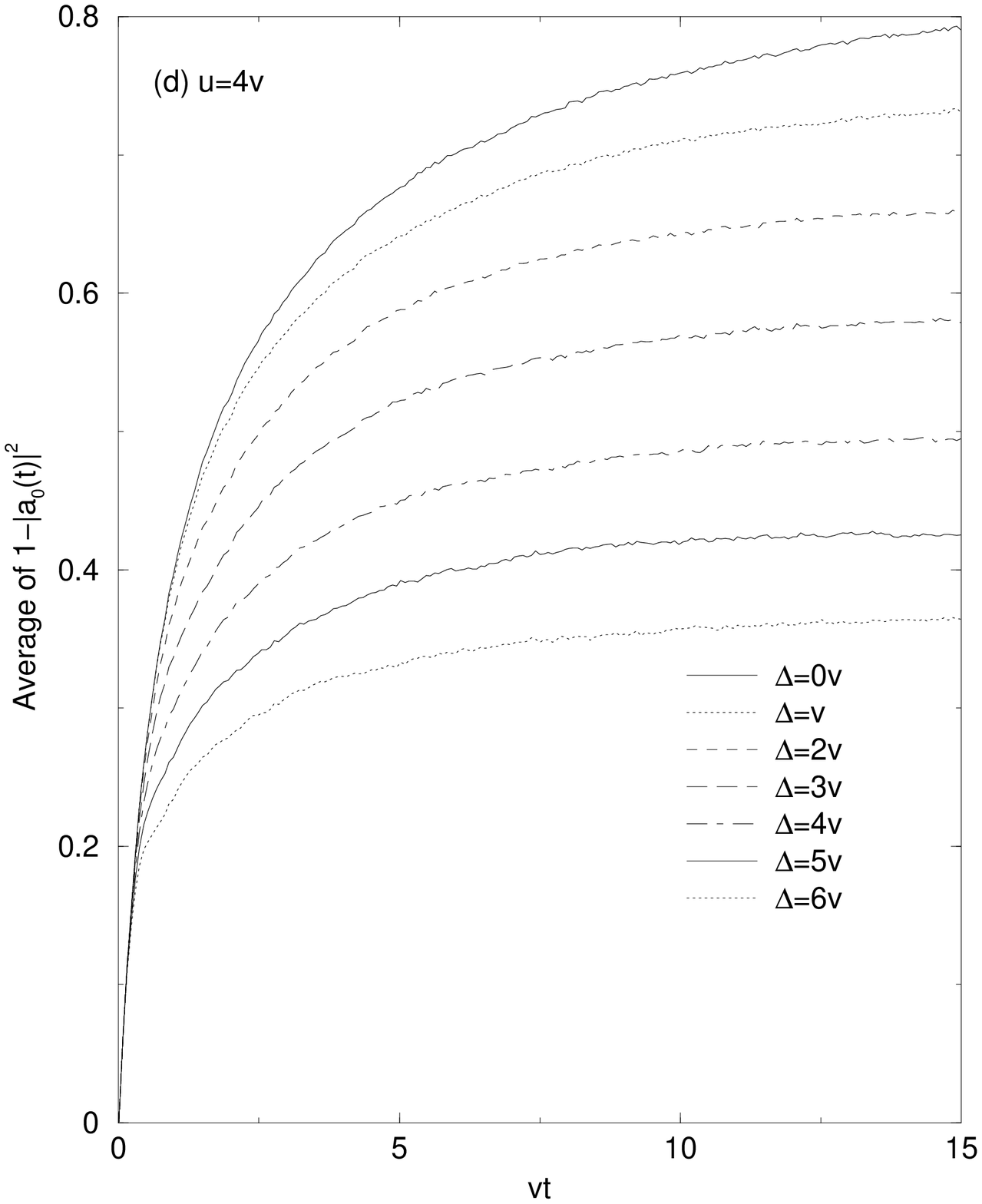}}
\end{center}
\caption{ Plots of $1-\left\langle\left|a_{0}\left(
t\right)\right|^{2}\right\rangle$ for $u=0$, $u=0.25v$, $u=v$, and
$u=4v$ for several values of the detuning $\Delta$, as shown on
each graph. For $u=0$, the exact result is shown for comparison.}
\label{azero2}
\end{figure}

\begin{figure}[tbp]
\begin{center}
\hbox{\epsffile{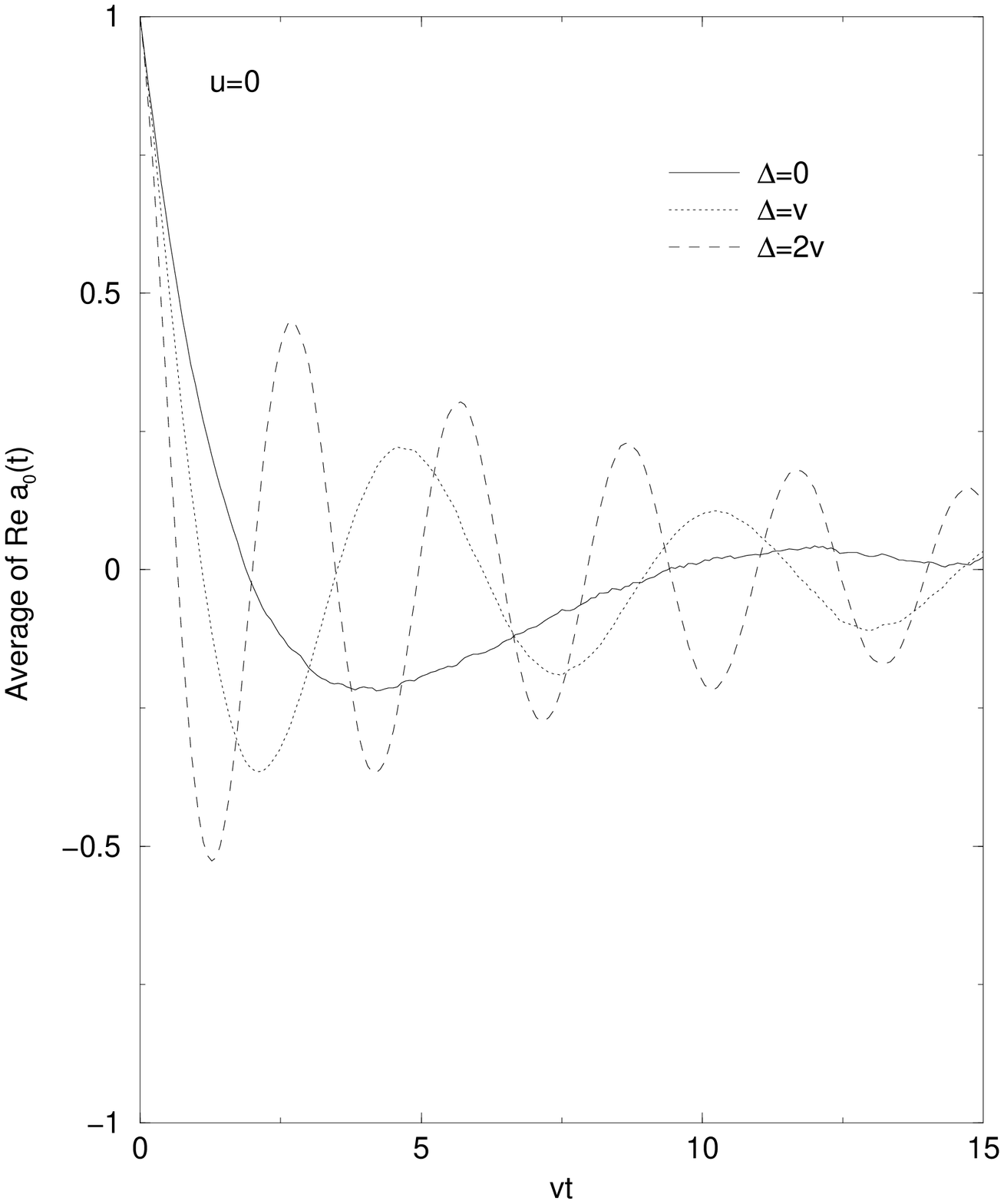}}
\pagebreak
\hbox{\epsffile{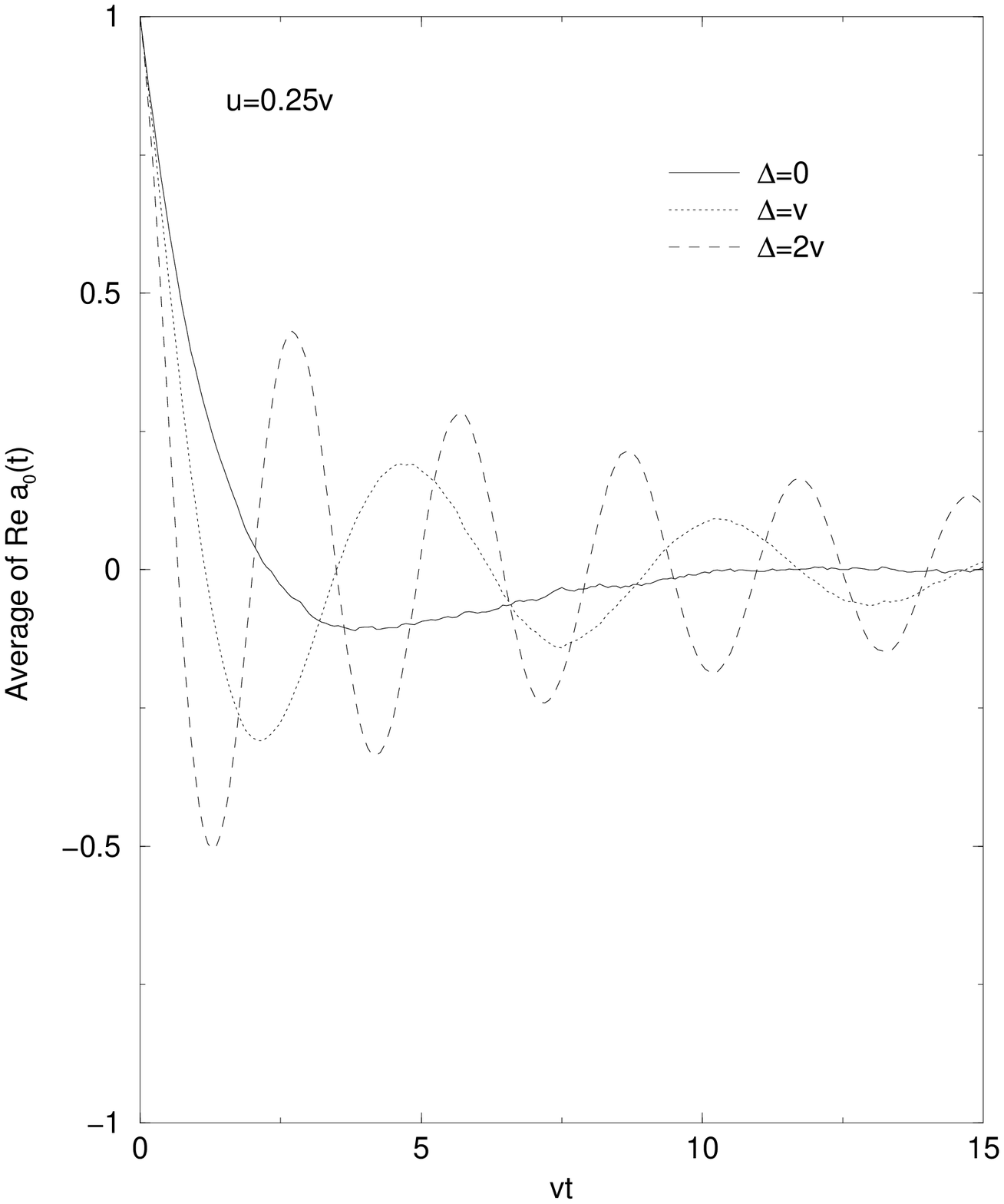}}
\pagebreak
\hbox{\epsffile{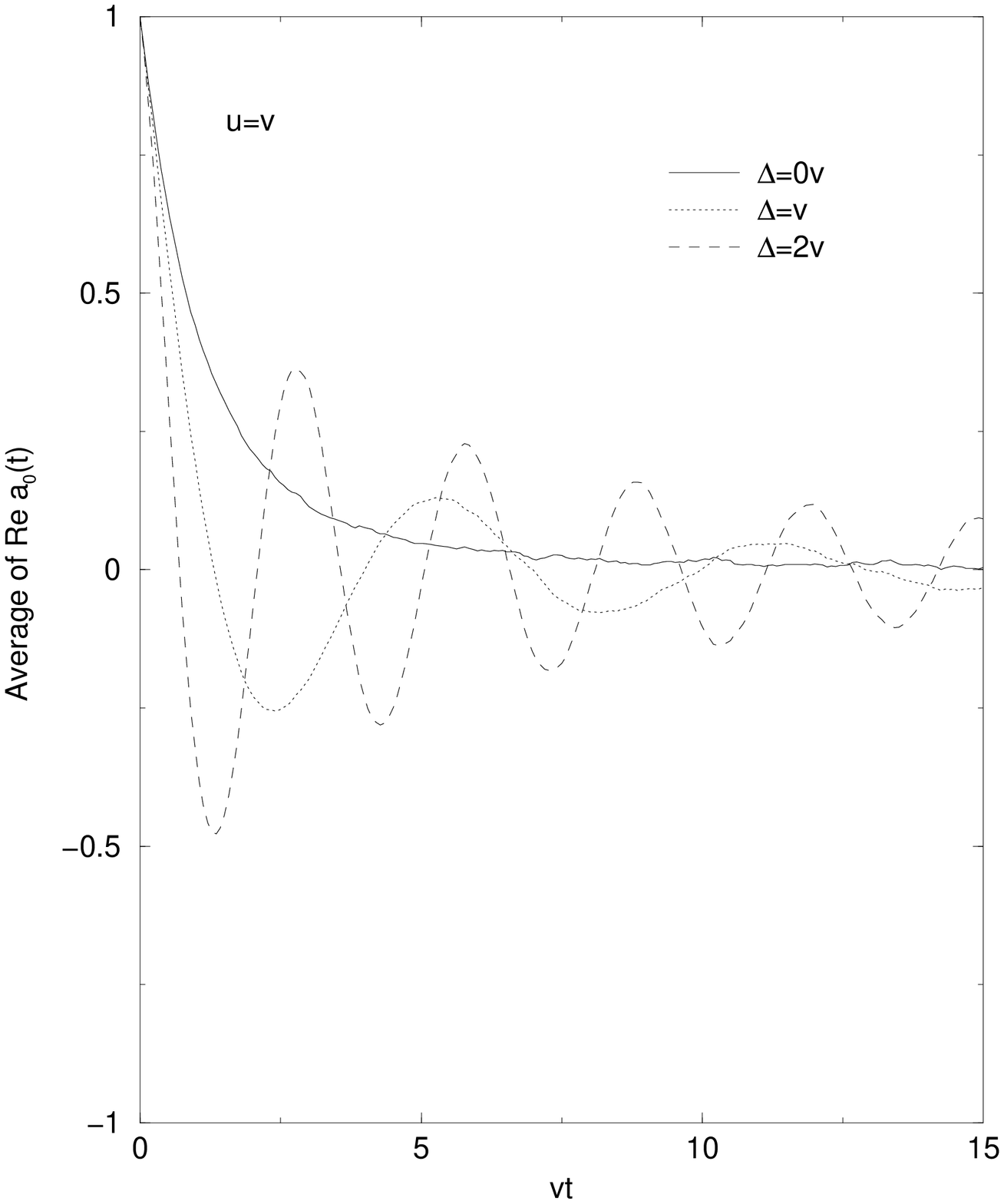}}
\pagebreak
\hbox{\epsffile{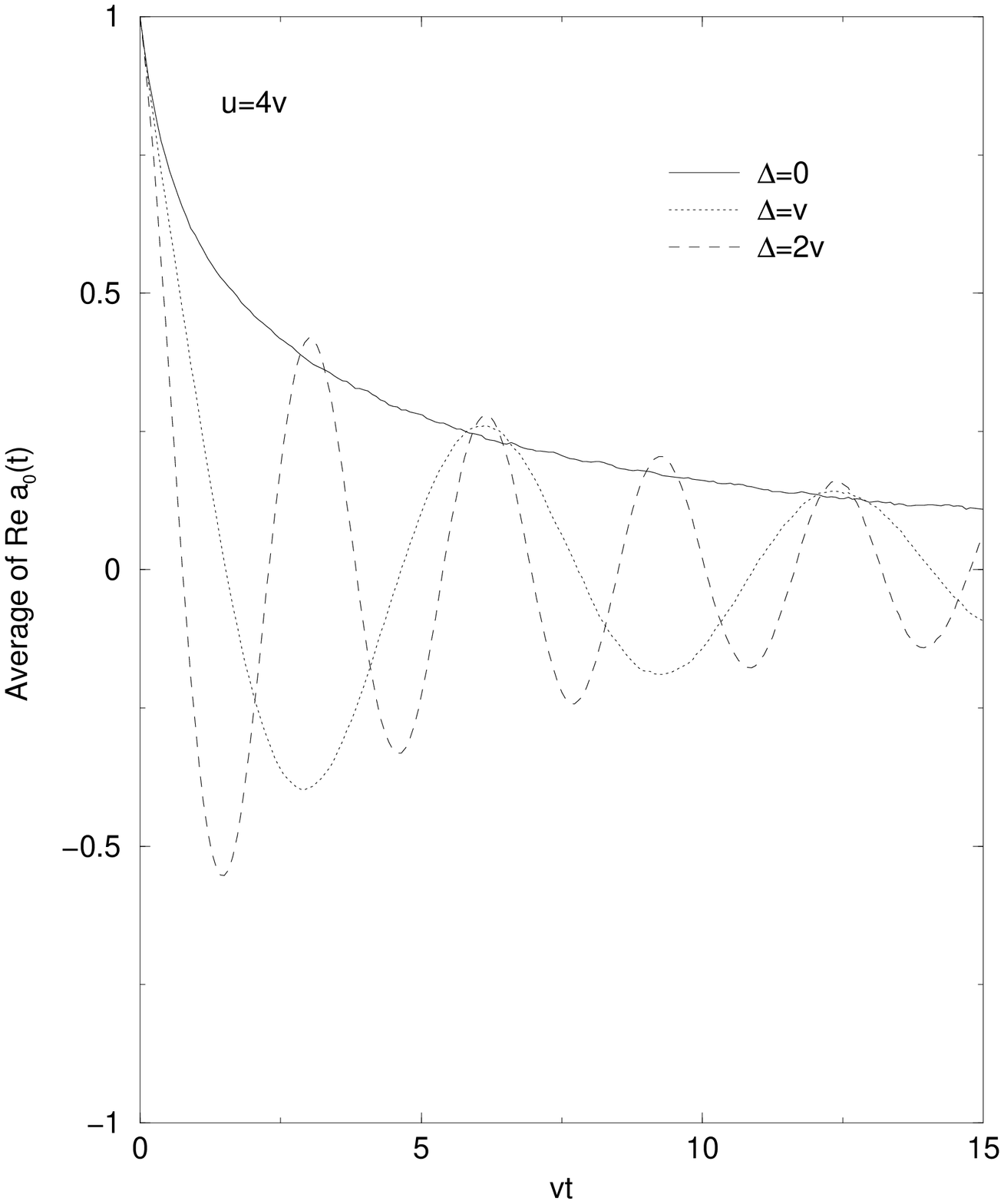}}
\end{center}
\caption{ Plots of ${\mathrm{Re}}\left\langle a_{0}\left(
t\right)\right\rangle$ for $u=0$, $u=0.25v$, $u=v$, and $u=4v$ for
three values of the detuning $\Delta$.} \label{azero_real}
\end{figure}

\begin{figure}[tbp]
\begin{center}
\hbox{\epsffile{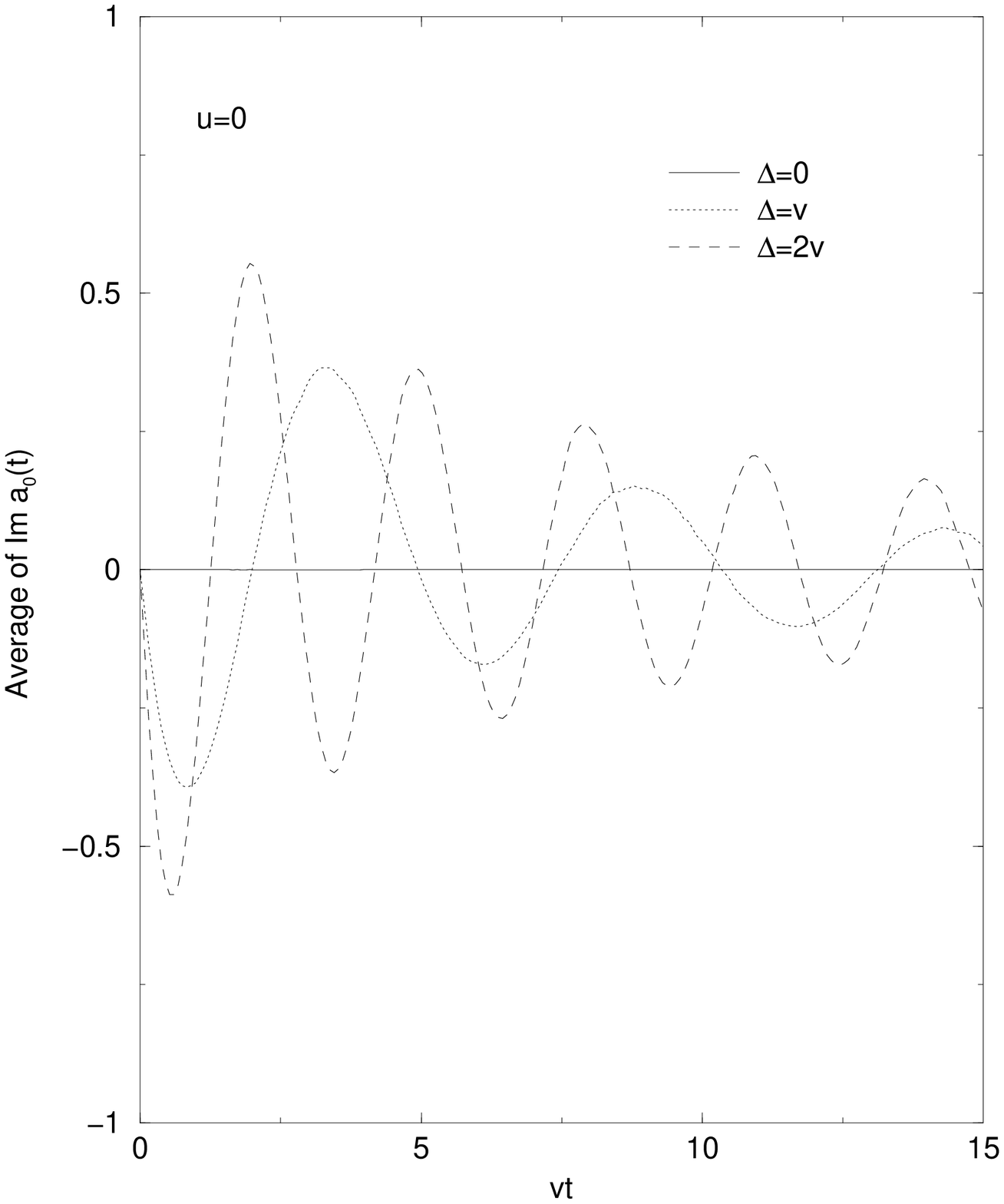}}
\pagebreak
\hbox{\epsffile{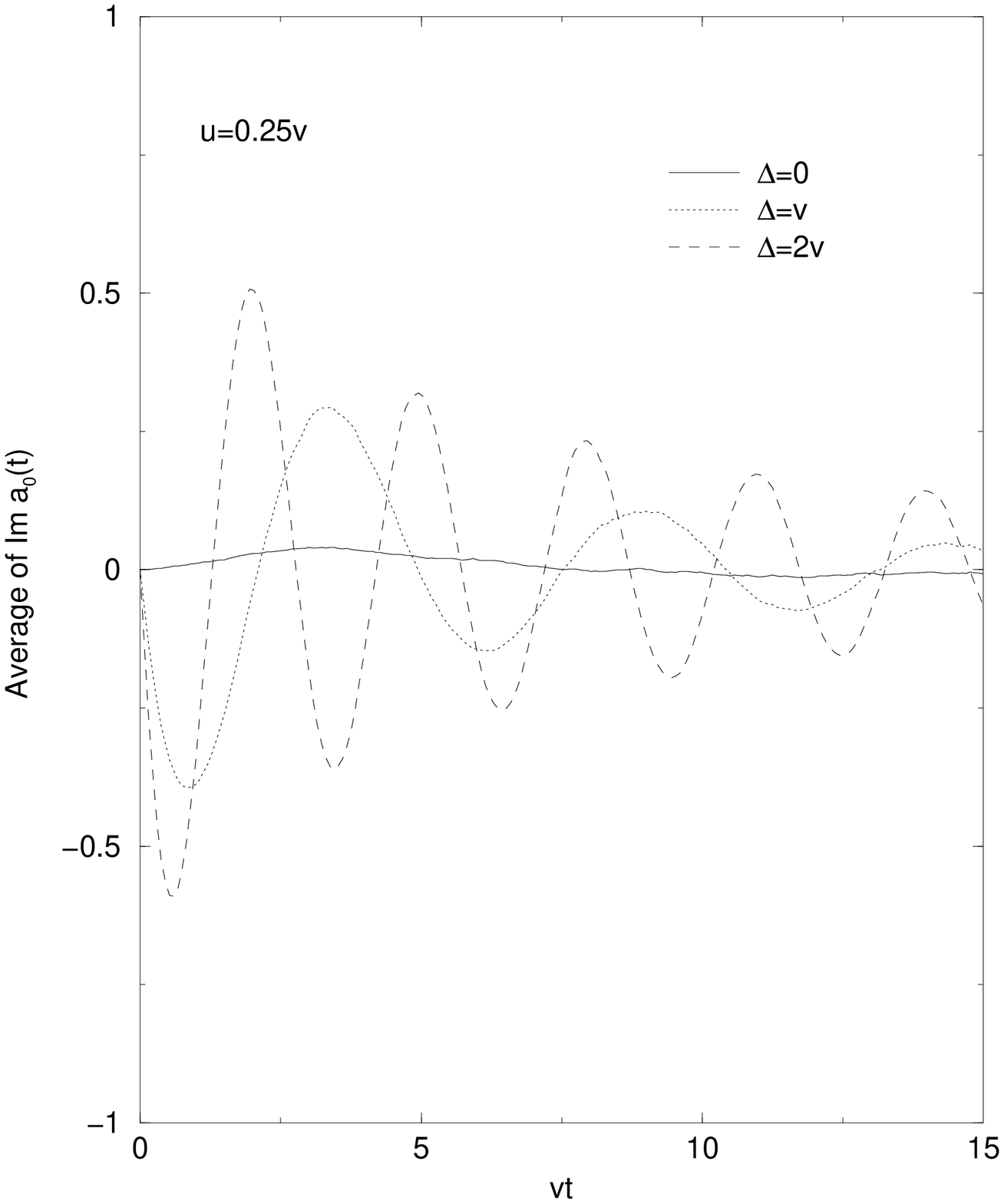}}
\pagebreak
\hbox{\epsffile{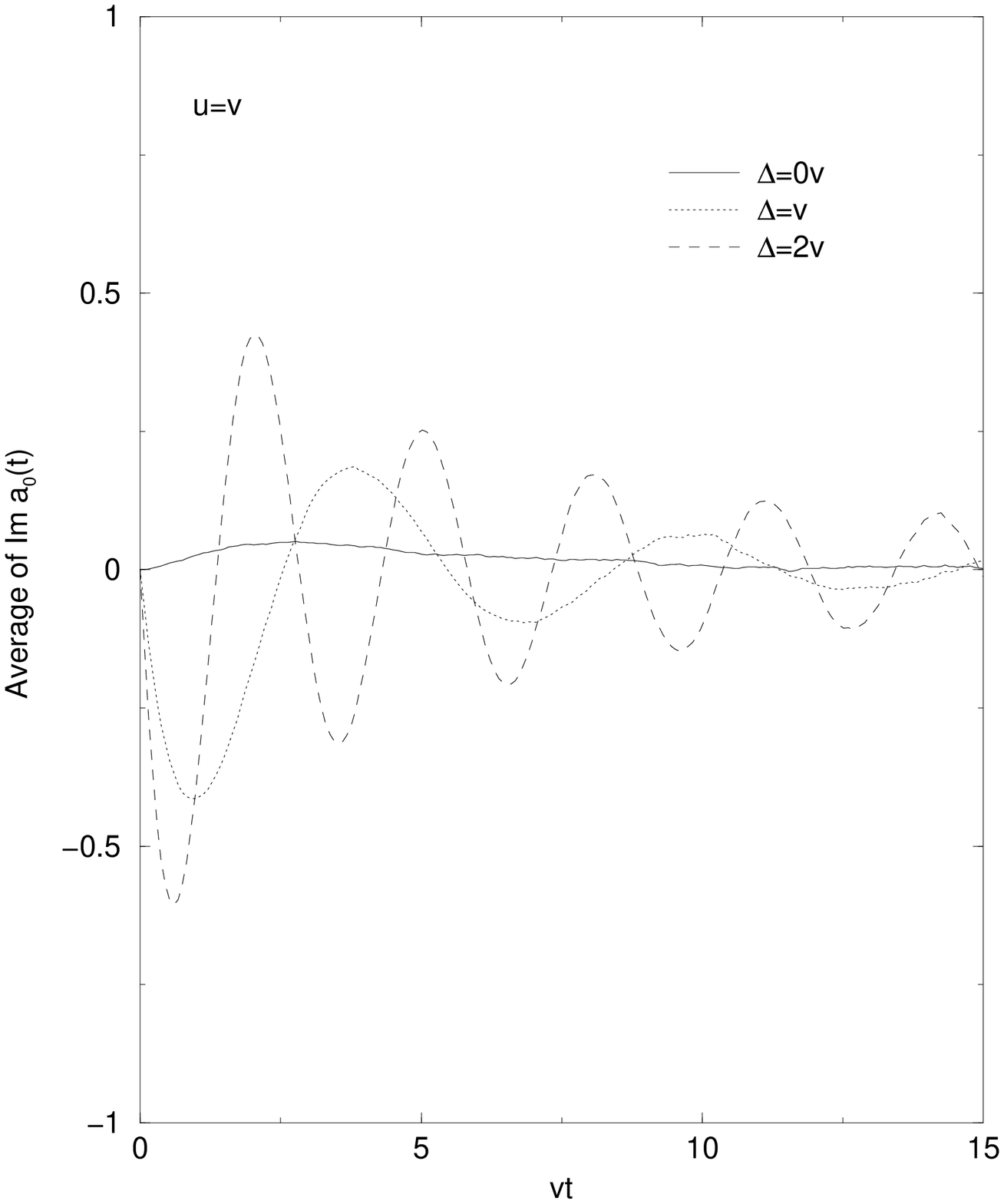}}
\pagebreak
\hbox{\epsffile{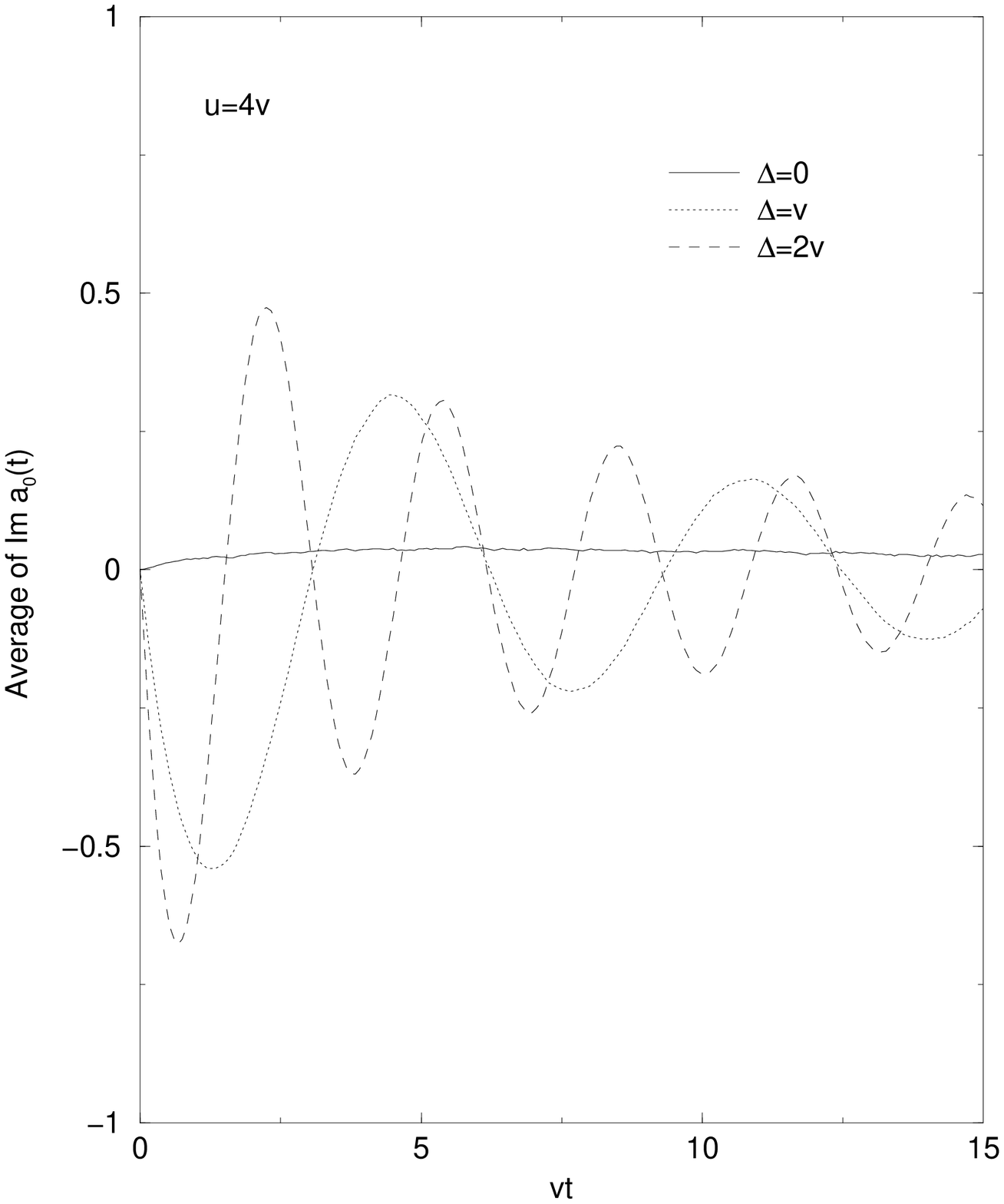}}
\end{center}
\caption{ Plots of ${\mathrm{Im}}\left\langle a_{0}\left(
t\right)\right\rangle$, for the same values of $u$ and $\Delta$ as
in Fig.~\ref{azero_real}. } \label{azero_imag}
\end{figure}

\begin{figure}[tbp]
\begin{center}
\hbox{\epsffile{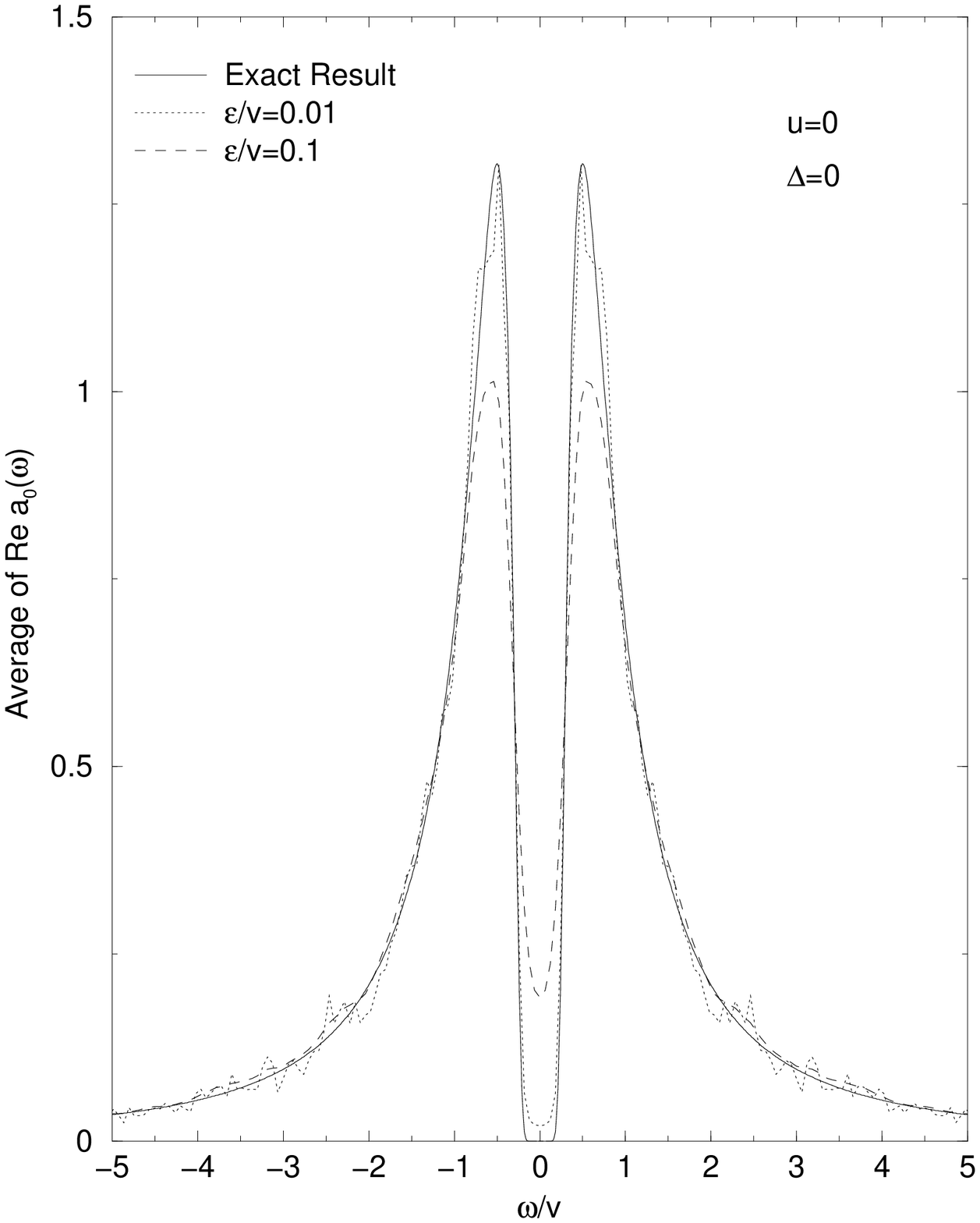}}
\pagebreak
\hbox{\epsffile{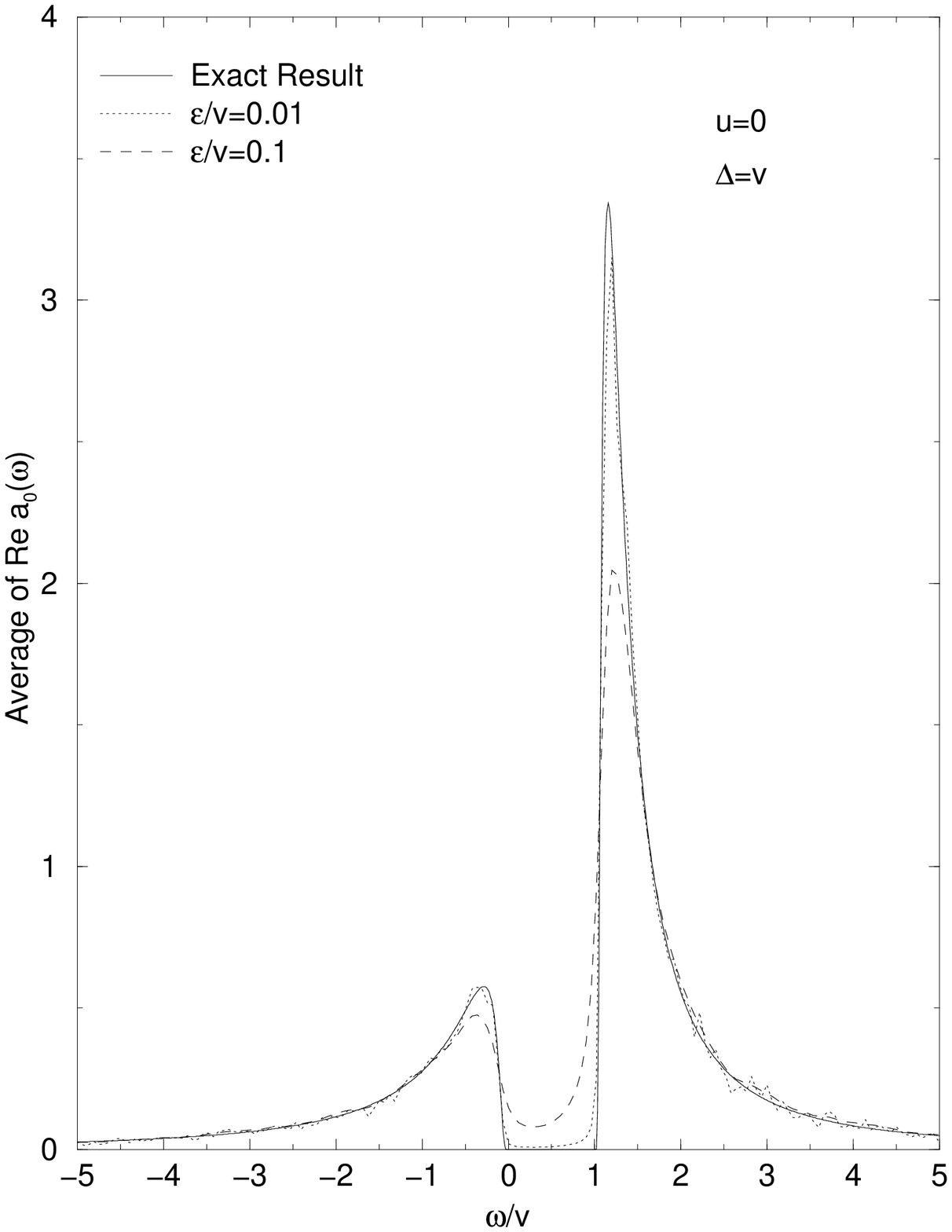}}
\pagebreak
\hbox{\epsffile{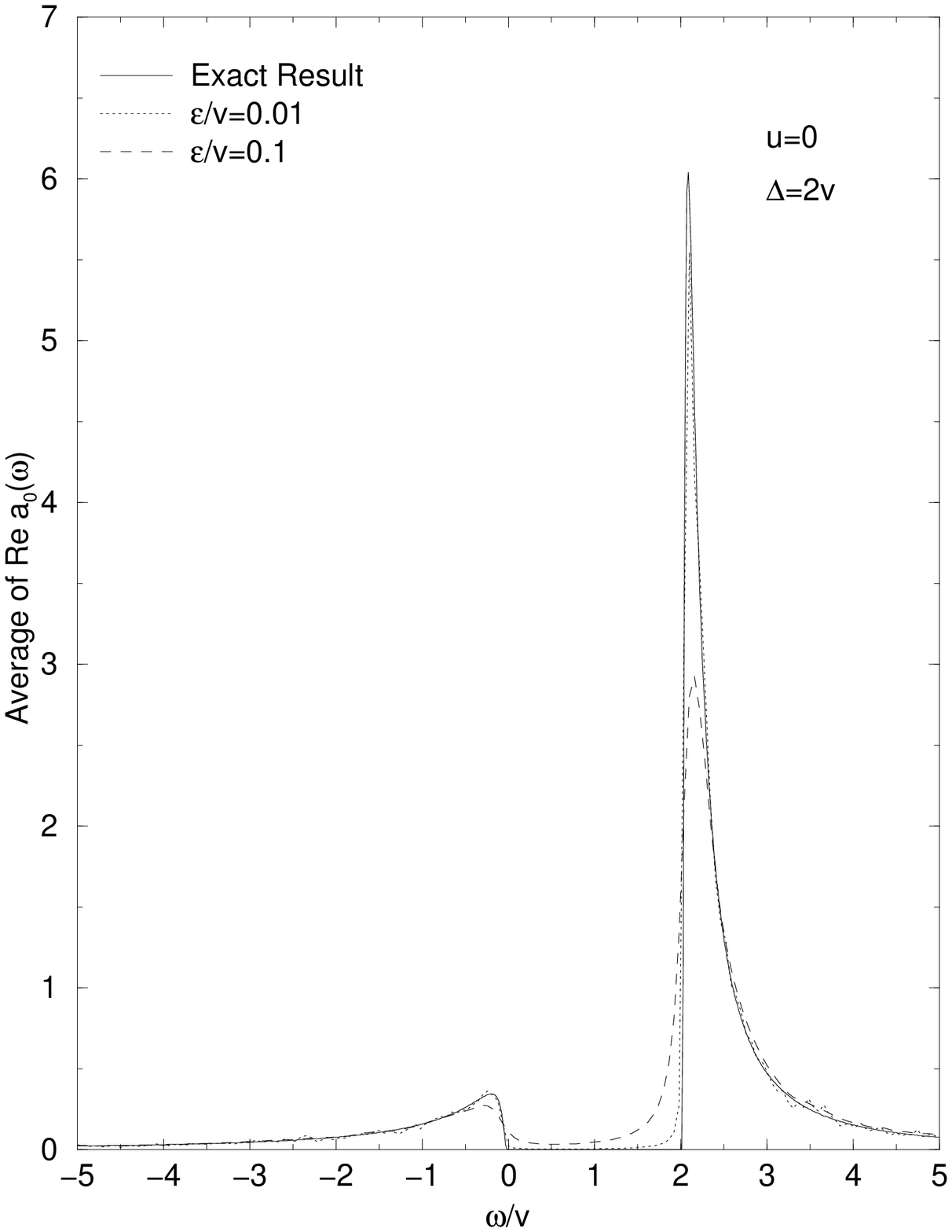}}
\end{center}
\caption{ Plots of ${\mathrm{Re}}\left\langle a_{0}\left(
\omega\right)\right\rangle$ for $u=0$ and $\Delta = 0$, $\Delta =
v$, and $\Delta = 2v$. Exact results are compared with numerical
simulations of ${\mathrm{Re}}\left\langle a_{0}\left(\omega +
i\varepsilon\right)\right\rangle$ for two values of
$\varepsilon$.} \label{u=0_azero_fourier}
\end{figure}

\begin{figure}[tbp]
\begin{center}
\hbox{\epsffile{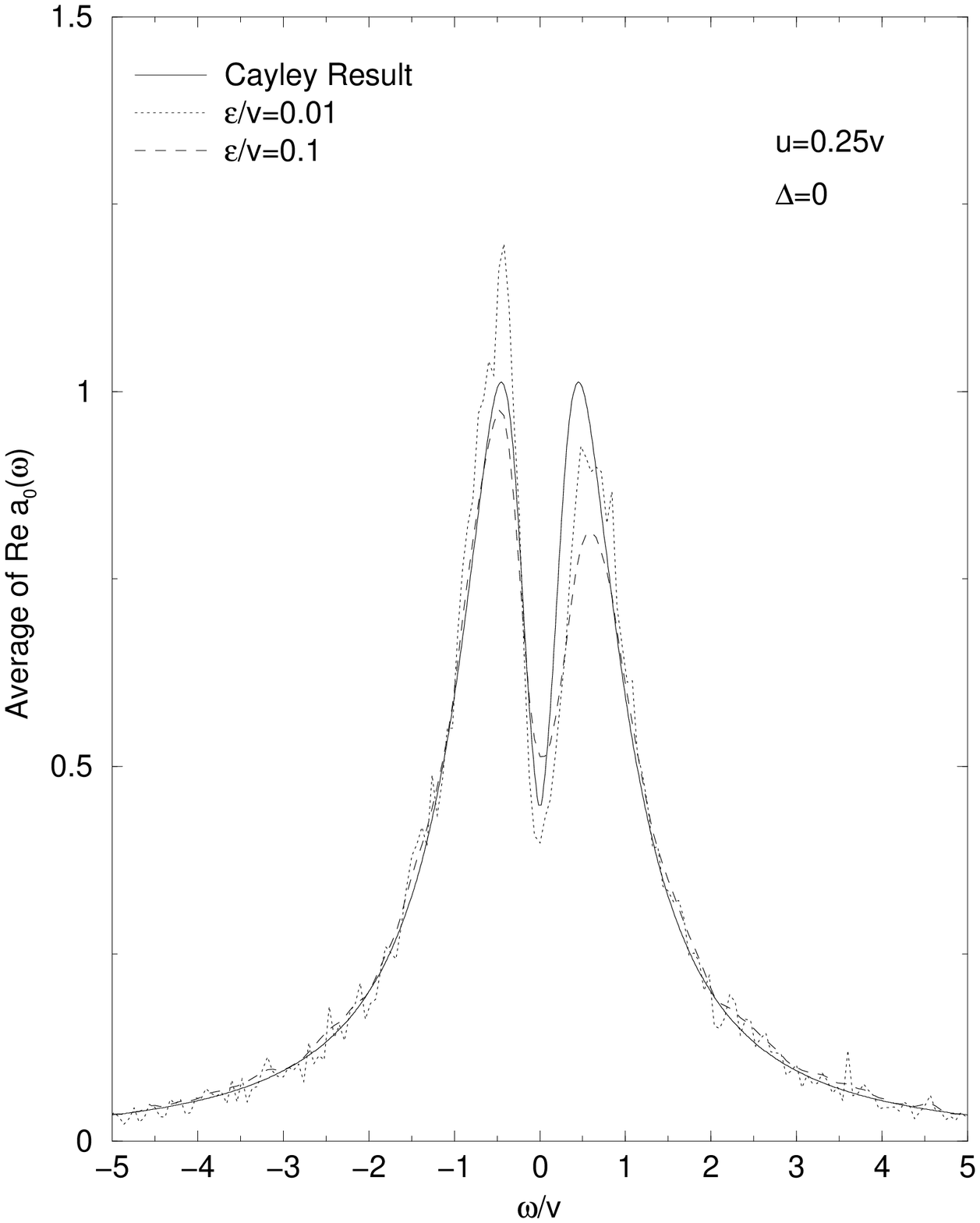}}
\pagebreak
\hbox{\epsffile{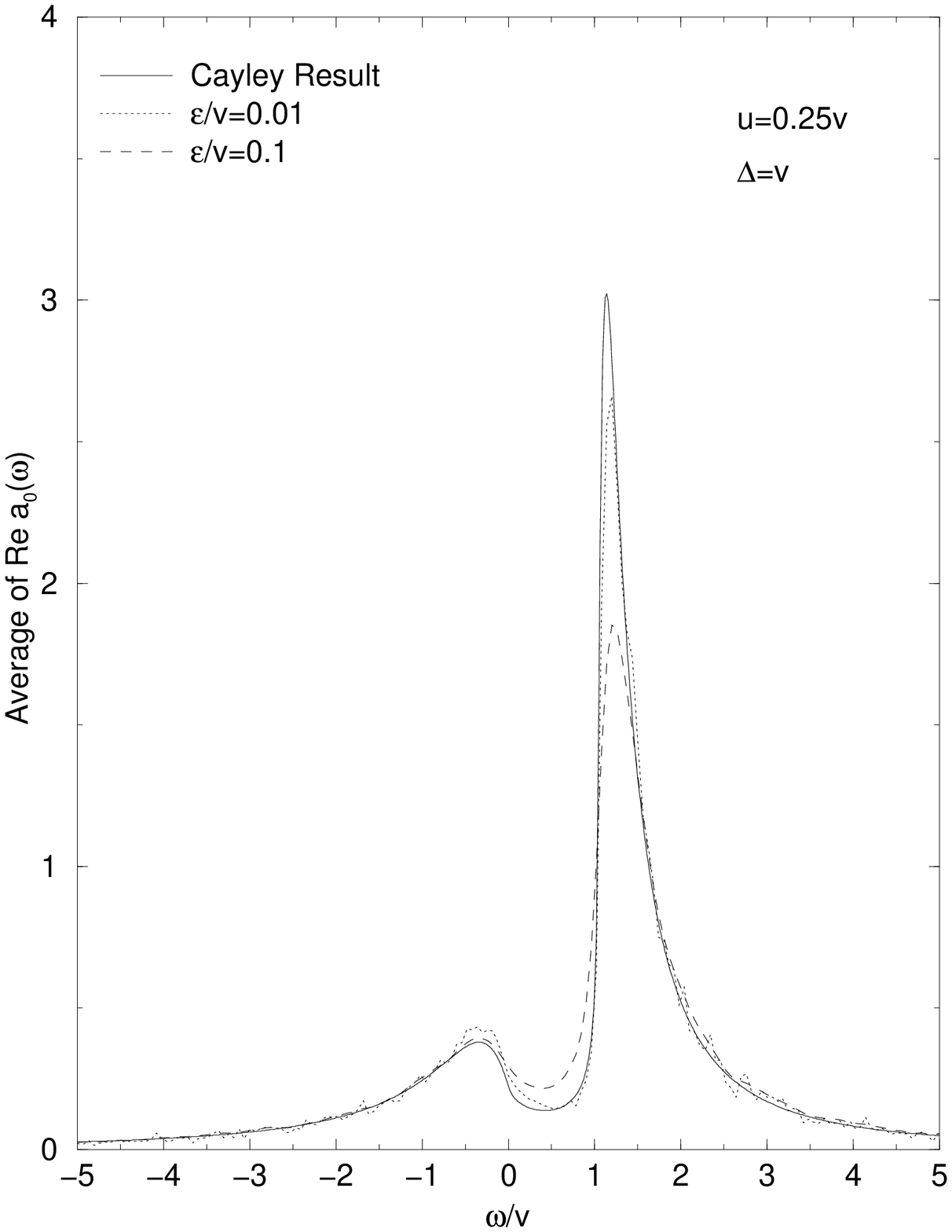}}
\pagebreak
\hbox{\epsffile{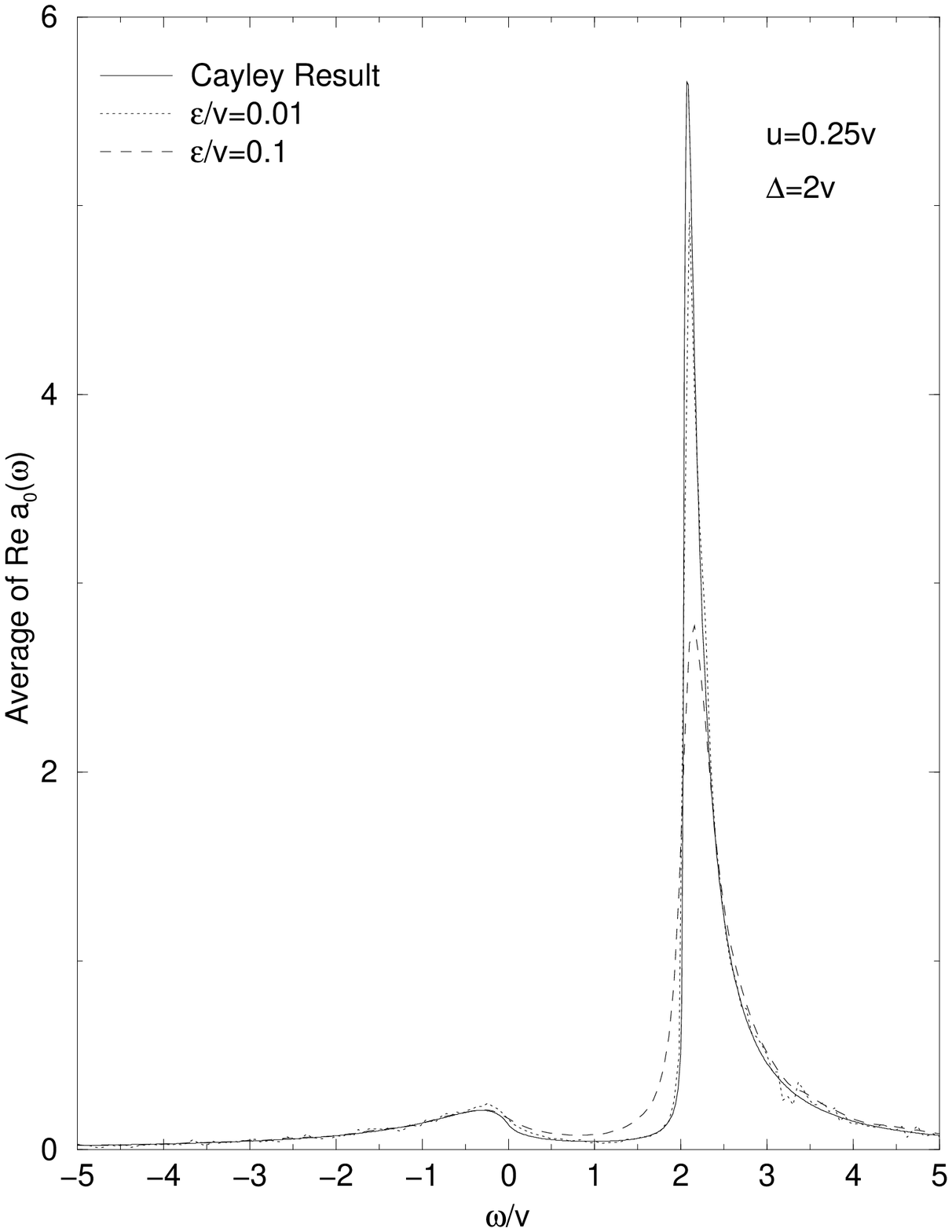}}
\end{center}
\caption{ Plots of ${\mathrm{Re}}\left\langle a_{0}\left(
\omega\right)\right\rangle$ for $u=0.25v$ and $\Delta = 0$,
$\Delta = v$, and $\Delta = 2v$.} \label{u=0.25v_azero_fourier}
\end{figure}

\begin{figure}[tbp]
\begin{center}
\hbox{\epsffile{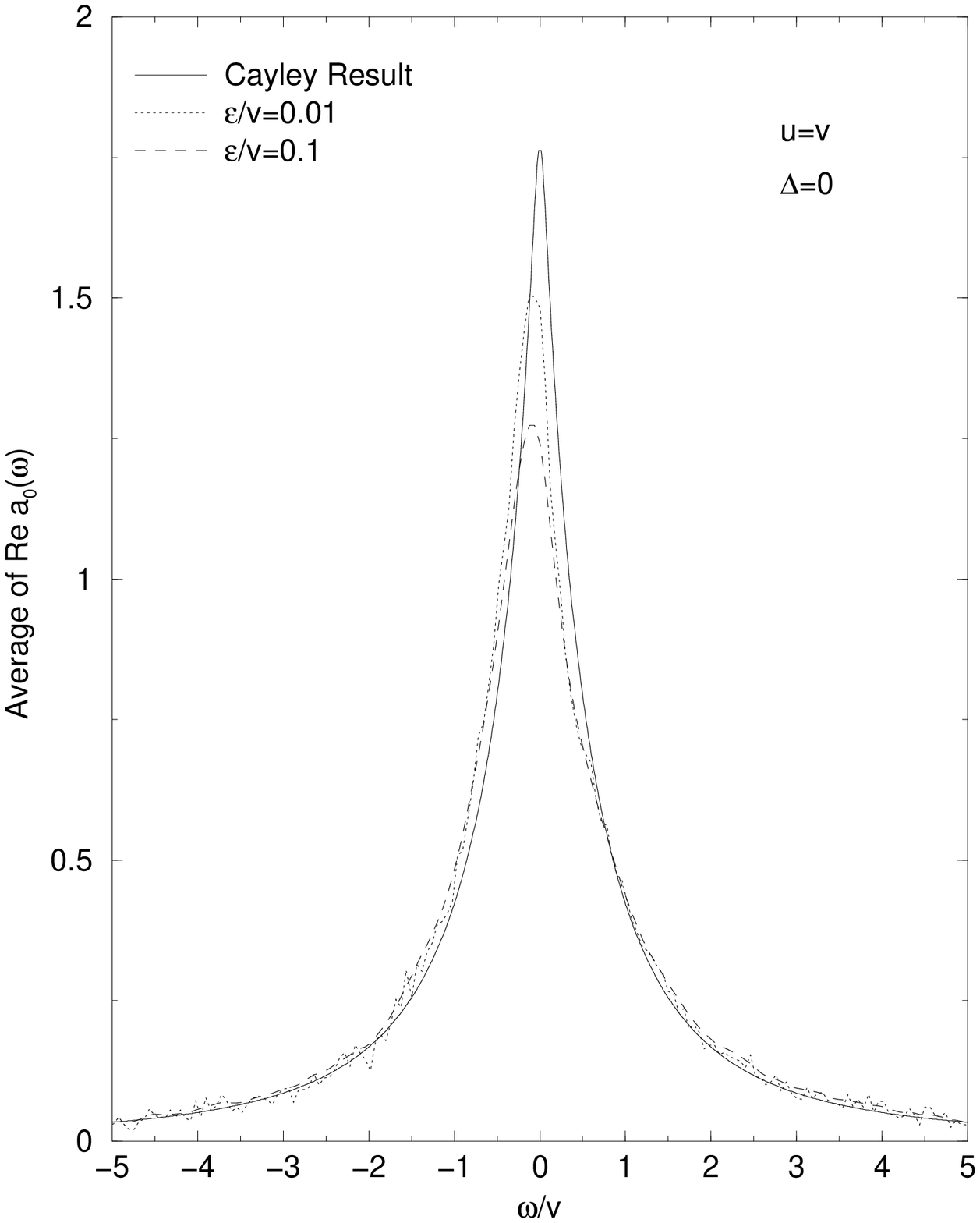}}
\pagebreak
\hbox{\epsffile{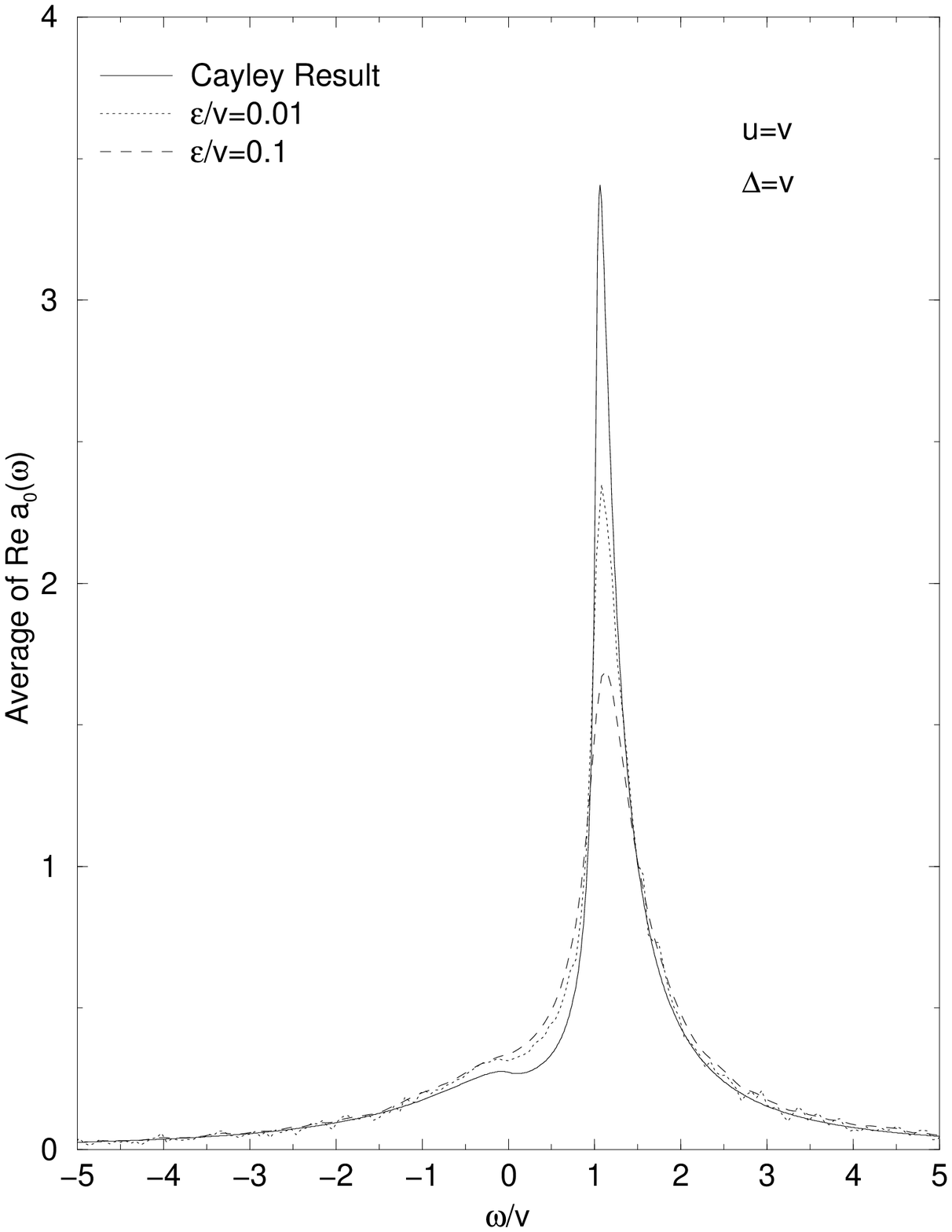}}
\pagebreak
\hbox{\epsffile{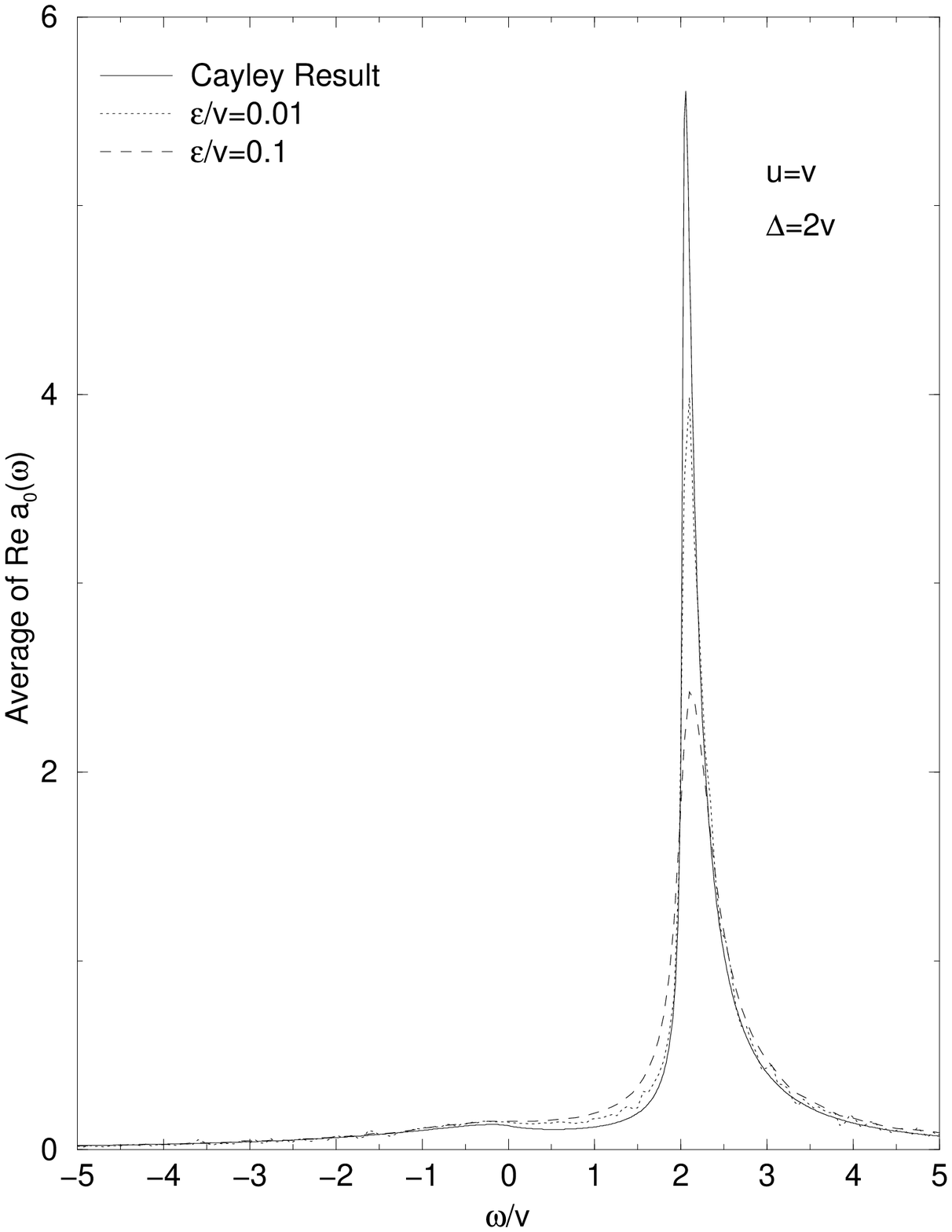}}
\end{center}
\caption{ As in Fig.~\ref{u=0.25v_azero_fourier}, but for $u=v$. }
\label{u=v_azero_fourier}
\end{figure}

\begin{figure}[tbp]
\begin{center}
\hbox{\epsffile{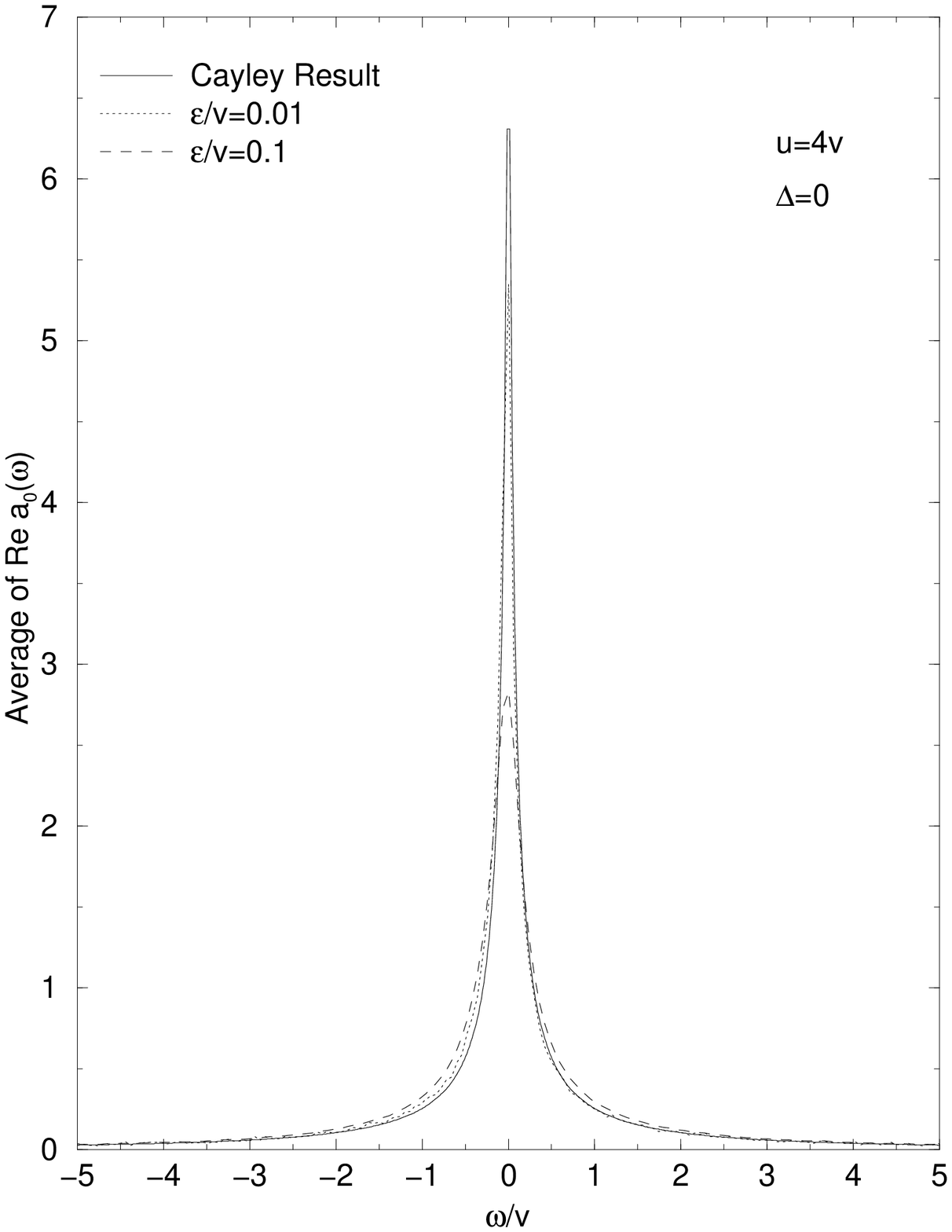}}
\pagebreak
\hbox{\epsffile{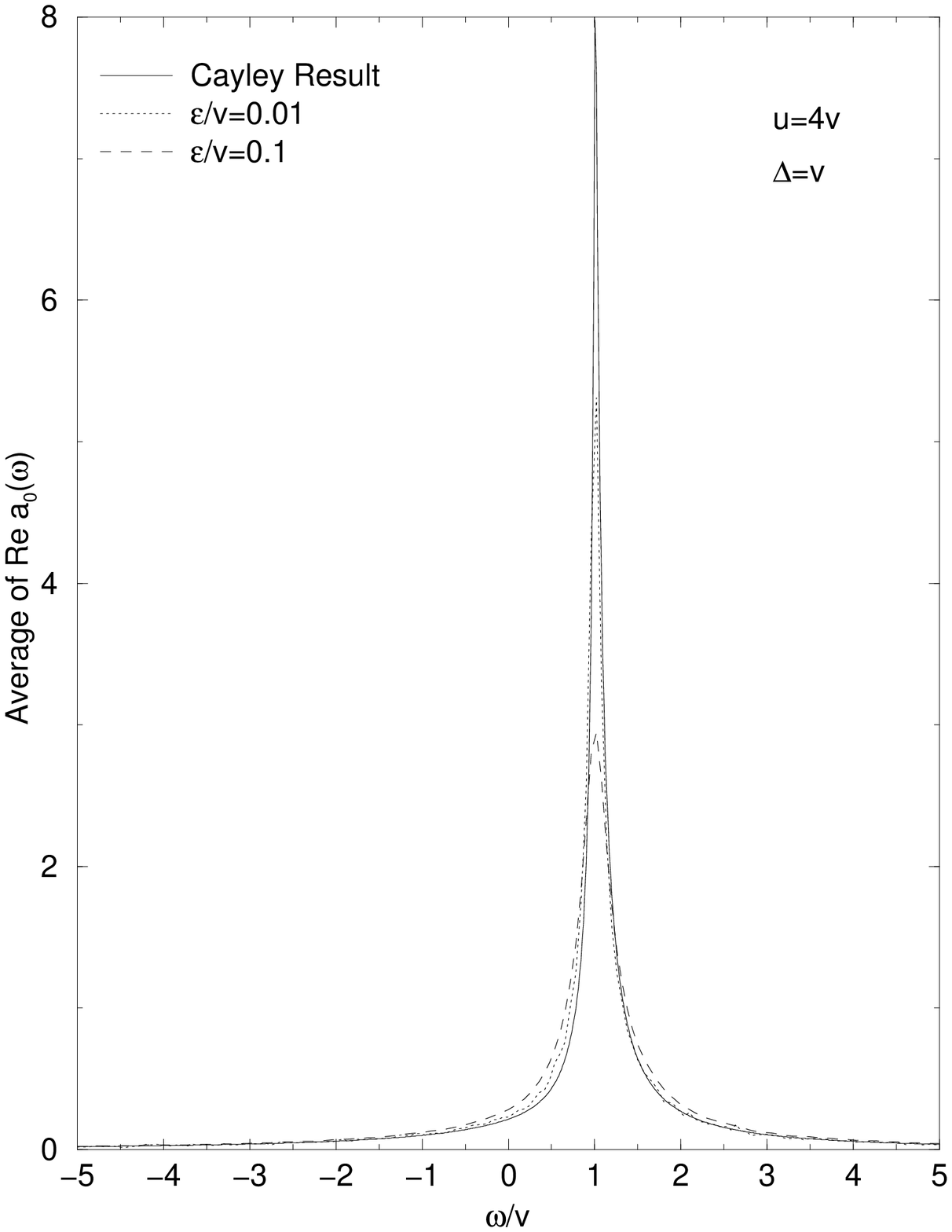}}
\pagebreak
\hbox{\epsffile{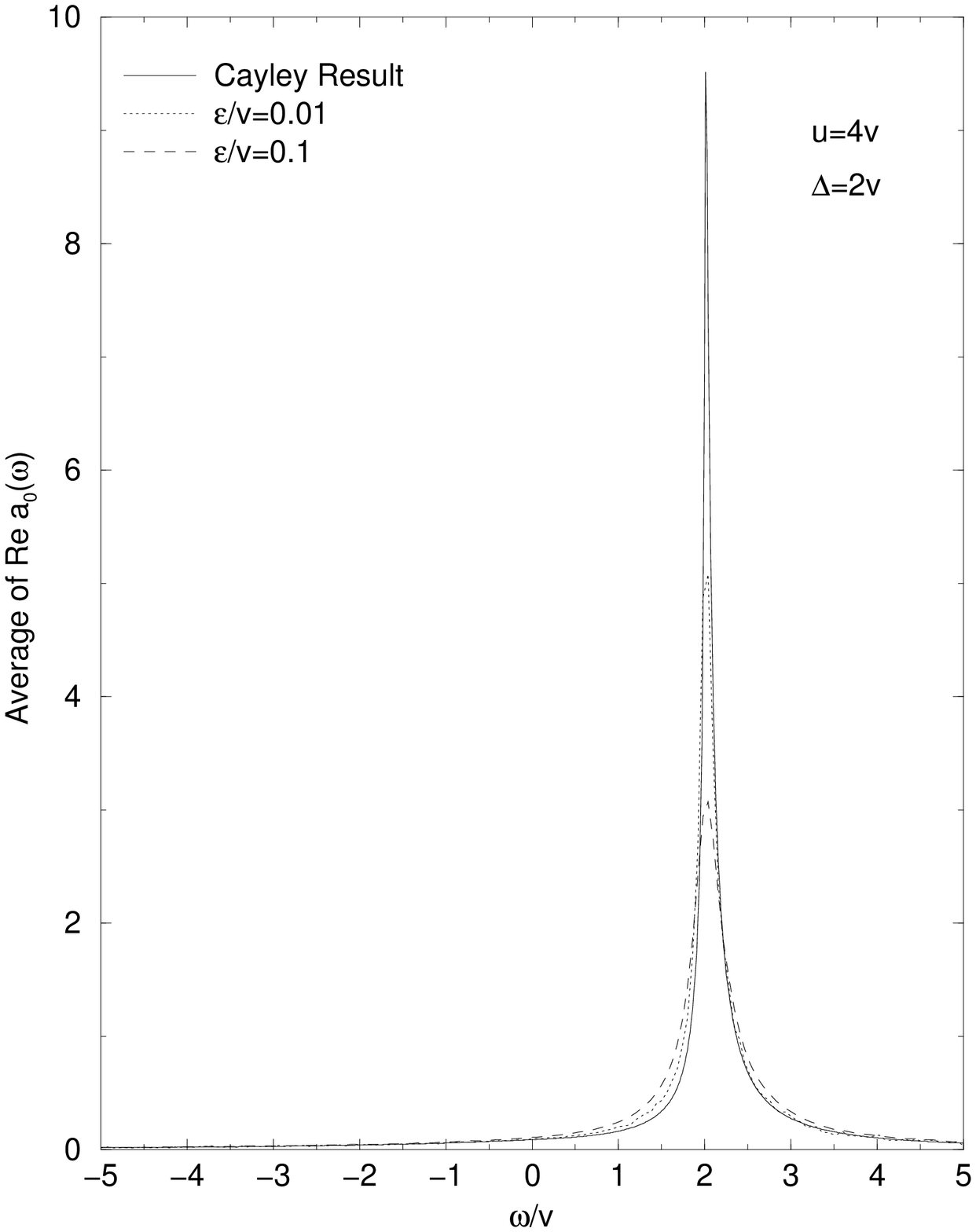}}
\end{center}
\caption{ As in Fig.~\ref{u=0.25v_azero_fourier}, but for $u=4v$.}
\label{u=4v_azero_fourier}
\end{figure}

The time-dependent resonance lineshape is obtained from the
ordinates of the curves in Fig.~\ref{azero2} at a given $vt$. In
general the lineshape is not Lorentzian, but we never find a split
line with a minimum at resonance as reported in
Ref.~\cite{mourachko}. Thus it is fair to characterize the line by
its FWHM. This cannot be directly compared with an experimental
width, which is usually the FWHM of a Lorentzian fitted to a noisy
experimental line, but it is indicative of trends. For a more
accurate comparison with experiment, one should of course
convolute the calculated lineshape with the broadening from other
effects.

\begin{figure}[tbp]
\begin{center}
\hbox{\epsffile{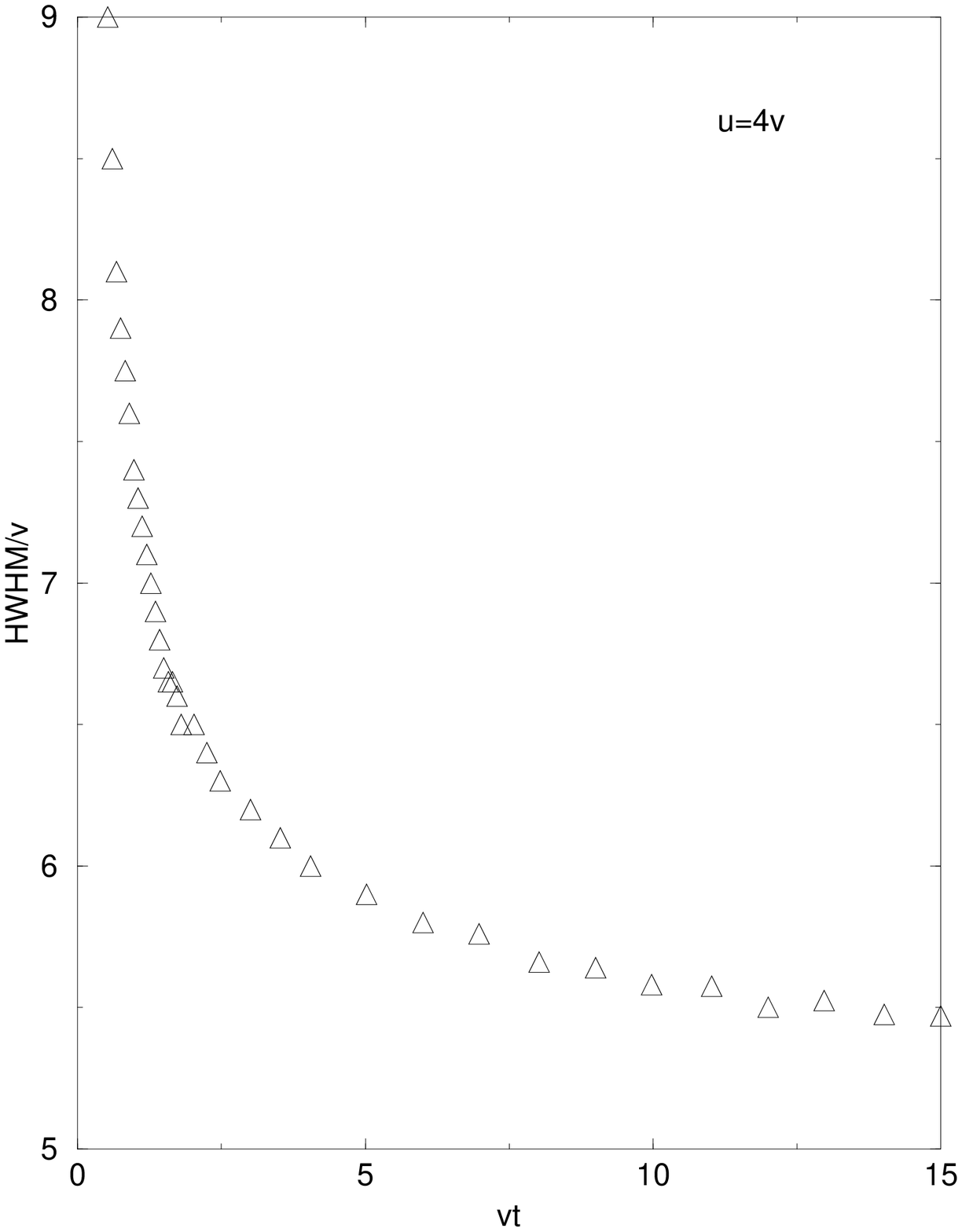}}
\end{center}
\caption{ A plot of the width as a function of time for $u=4v$.}
\label{widthvst}
\end{figure}

Fig.~\ref{widthvst} shows one half of the FWHM, $w$, as a function
of $vt$ for $u=4v$. One sees the typical decrease of $w$ from a
universal $t^{-1}$ behavior at small $vt$ towards an asymptotic
value, which is almost reached at $vt=15$. The $t^{-1}$ law, which
results simply from the uncertainty principle, has already been
discussed in Ref.~\cite{fcb}, where on the basis of a small
$\Delta$ expansion it was estimated that $w=\sqrt{12}/t$ for small
$t$.

\begin{figure}[tbp]
\begin{center}
\hbox{\epsffile{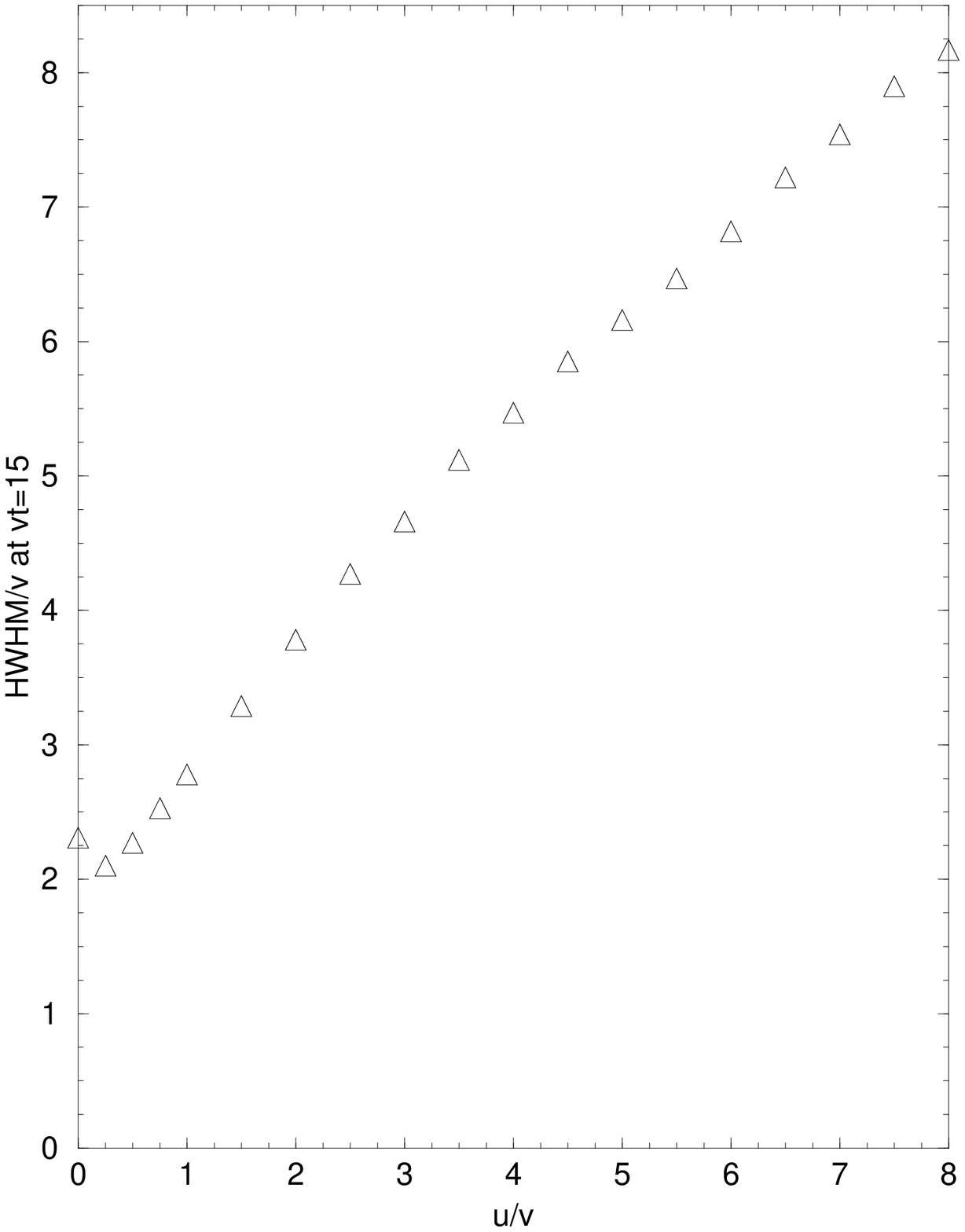}}
\end{center}
\caption{ A plot of the width at $vt=15$ as a function of the
ratio $u/v$.} \label{widthratio}
\end{figure}

The near-asymptotic value of $w$ at $vt=15$ is shown in
Fig.~\ref{widthratio} as a function of $u/v$. Overall, the graph
is roughly linear except at small $u/v$, although a careful
examination shows that there are really two linear regimes with
somewhat different slopes that meet at $u/v \approx 3.5$. There is
an unexpected minimum around $u=v/4$, which is not a numerical
artifact, but can be seen directly from Fig.~\ref{azero2} by
comparing the plot for $u=v/4$ with those for $u=0$ and $u=v$.
What happens is that the resonance line always grows taller and
wider as $u$ increases, but for small $u$ its height (the signal
at $\Delta=0$) grows faster than its width.

\section{CONCLUSIONS}
\label{discussion}

In this paper, we report results of ongoing work on the general
question: what are the electrodynamics and the optics of a gas of
excited atoms interacting via dipolar fluctuations, for times
short compared to the atom-atom collision time? We are considering
in particular the time evolution of a resonant $ss^{\prime
}\rightarrow pp^{\prime }$ process when $N$ atoms in state $s$ and
$N^{\prime }$ in state $s^{\prime}$ are initially present, and all
atoms are randomly distributed. At least one of these two initial
states must be an excited state for the process to occur
resonantly. Other resonant processes can be mapped on this
prototype. For instance, the process $pp\rightarrow ss^{\prime }$
discussed in Refs.~\cite{akulin} and \cite{mourachko} is handled
by setting $s^{\prime }=s$ and reversing the roles of $s$ and $p$.
If the atoms are in Rydberg states then they have known energies
and dipole moment matrix elements and the dipole-dipole
interaction is dominant, so there are no adjustable parameters in
the theory. More importantly for the experiments, with a frozen
gas of Rydberg atoms one can easily do time-resolved spectroscopy
because the evolution time scale is measured in microseconds.
However, for the theory it does not matter that the atoms are of
Rb or Cs, since other atoms in excited states would behave
similarly on the corresponding time scale. In the language of the
solid-state and chemical physics literatures, our problem can be
described as resonant energy transfer to Frenkel excitons.

We have carried out detailed and extensive calculations, truly
``computer experiments'', for a resonant $ss^{\prime}\rightarrow
pp^{\prime }$ process when a single $s^{\prime}$ atom is initially
present in a frozen gas of $s$ atoms. Neglecting the propagation
of the $s^{\prime}p^{\prime}$ exciton to keep things simple, we
consider in effect the resonant excitation and propagation of an
$sp$ exciton. The results apply, in practice, whenever only a few
excitons of a given type are created, so that exciton-exciton
interactions are negligible.

We can compute anything we wish about this system, but in this
paper we report representative results for two quantities only:
the average probability amplitude that no exciton has been
created, $\left\langle a_{0}\right\rangle$, and the average
probability that an exciton has been created, $1-\left\langle
\left| a_{0}\right|^{2}\right\rangle$. One striking result is that
the initial transition rate for this process is independent of
exciton propagation and of detuning from resonance for
surprisingly long times, and can thus be explicitly evaluated.
Another notable feature is that the transition amplitude at
detuning $\Delta$ shows persistent oscillations with period
$\Delta$, in particular when exciton propagation is a strong
effect, as for $u=4v$. Correspondingly, the spectral density for
exciton creation is strongly peaked at $\Delta$, and becomes
narrower as $u$ increases. Thus, the universal phenomenon of line
narrowing actually simplifies the problem in the experimentally
interesting limit of strong coupling ($u\gg v$).

With a wealth of numerical results at our disposal, we have
attempted to develop approximate formulas that are not simply fits
to a particular numerical result. We have obtained good
approximations for $\left\langle a_{0}\right\rangle$, but not yet
for $\left\langle \left| a_{0}\right|^{2}\right\rangle$, which is
notoriously more difficult to obtain than $\left\langle
a_{0}\right\rangle$ and quite different from $\left|\left\langle
a_{0}\right\rangle\right|^{2}$. To develop approximation schemes
we have drawn, and will continue to draw, from the literature on
wave functions in random systems, and in particular on earlier
work on excitons in disordered solids and liquids\cite{lw}. The
best results, in our case, are given by the Cayley tree
approximation.

Most of the work on random systems deals with on-site disorder or
short-range interactions. On the contrary, the dipolar interaction
$U\left(r\right)\sim 1/r^{3}$ is long range and has the
distribution $P\left(U\right)\sim 1/U^{2}$ for a random
distribution of $r$. The distribution of the cumulative variable
${\mathcal{U}}$, where
${\mathcal{U}}^{2}=\sum_{k}U^{2}\left({\mathbf{r}}_{k}\right)$,
plays a major role in the theory. It is given by
\begin{equation}
P\left({\mathcal{U}}\right)=\frac{1}{\sqrt{\pi
}}\frac{u}{{\mathcal{U}}^{2}}\exp\left(
-\frac{u^{2}}{4{\mathcal{U}}^{2}}\right), \label{P}
\end{equation}
where $u$ is given by the appropriate equivalent of
Eq.~(\ref{u}).\footnote{The angular dependence of $U$ modifies the
value of $u$, leaving unchanged the functional form of
Eq.~(\ref{P}). See Ref.~\cite{fcb}, Appendix A.} Note that
$P\left({\mathcal{U}}\right)$ is not Gaussian and the central
limit theorem does not apply, because $P\left(U\right)$ does not
have an r.m.s.\ value and is not even normalizable. Therefore it
is not surprising that, as shown by the simulations, the
distribution of eigenvalues and other statistical properties of
the matrix $U_{kl}$ are very different from those of a standard
random matrix. This implies that random system results obtainable
by random matrix theory are not applicable to dipolar interactions
in an ideal gas. Luckily, the distribution in Eq.~(\ref{P}) makes
the Cayley tree equations for $\left\langle a_{0}\right\rangle$
exactly soluble in this case.

\acknowledgments Acknowledgment is made to the Thomas F. and Kate
Miller Jeffress Memorial Trust for the support of this research.
We also wish to thank Ziad Maassarani for many useful discussions.

\widetext

\end{document}